\documentclass{aa}  

\usepackage{graphicx}
\usepackage{txfonts}
\usepackage{natbib}
\bibpunct{(}{)}{;}{a}{}{,} 
\usepackage{booktabs}
\usepackage{multicol}
\usepackage{graphicx}
\usepackage{fancyhdr}
\usepackage[hidelinks,colorlinks=true,linkcolor=blue,citecolor=blue]{hyperref}
\usepackage{amsmath}	
\usepackage{amssymb}	
\usepackage[inter-unit-product=\cdot]{siunitx}
\usepackage{enumerate}
\usepackage{multirow}
\usepackage{mathtools}
\usepackage{comment}
\usepackage{longtable}
\usepackage{tabularx}
\usepackage{xcolor} 
\usepackage{ulem} 
\usepackage{comment}
\usepackage{cuted}
\usepackage{soul}

\newcommand{\Msun}{\,\mathrm{M}_\odot}

\newcommand{\MSMBH}{M_\bullet}
\newcommand{\tonde}[1]{\left(#1\right)}
\newcommand{\quadre}[1]{\left[#1\right]}

\newcommand{\new}[1]{\textcolor{red}{#1}}

\newcommand{\fastcluster}[1]{{\sc fastcluster}}
\newcommand{\sevn}[1]{{\sc SEvN}}
\newcommand{\pagn}[1]{{\fontfamily{lmtt}\selectfont pAGN}}
\newcommand{\tsunami}[1]{{\sc tsunami}}
\newcommand{\princess}[1]{Princess}
\newcommand{\mcfacts}[1]{{\sc McFacts}}

\newcommand{\orcidicon}[1]{\href{https://orcid.org/#1}{\includegraphics[width=11pt]{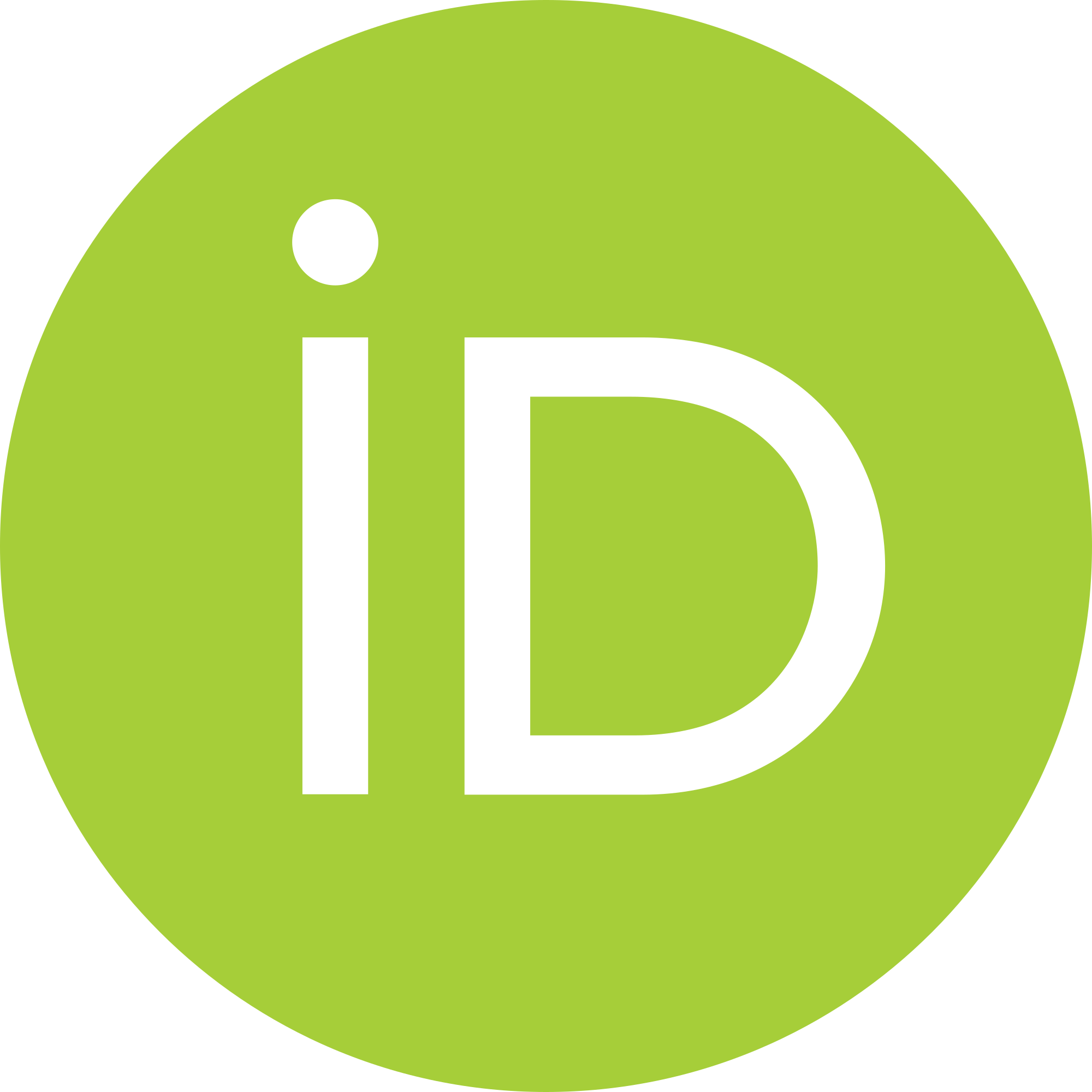}}}
\newcommand{\orcid}[1]{\href{https://orcid.org/#1}{\protect\orcidicon{#1}}}

\DeclareSIUnit\year{yr}
\DeclareSIUnit\au{AU}
\DeclareSIUnit\parsec{pc}
\DeclareSIUnit\erg{erg}

\defcitealias{SG}{SG}
\defcitealias{TQM}{TQM}
\defcitealias{Bellovary_2016}{B16}
\defcitealias{Grishin_2024}{G24}
\defcitealias{Kanagawa_2018}{K18}

\begin{document} 

  \title{The role of migration traps\\in the formation of binary black holes in AGN disks}

   \author{Maria Paola Vaccaro
   \thanks{e-mail: \href{mailto:mariapaolavaccaro@gmail.com}{mariapaolavaccaro@gmail.com}}\inst{1}\orcid{0000-0003-3776-9246},
            Yannick Seif\inst{1}\orcid{0009-0007-0937-5561},
          \and
          Michela Mapelli
          \inst{1,2,3,4}\orcid{0000-0001-8799-2548}}
    \authorrunning{M. P. Vaccaro et al.}
    \institute{
      $^{1}$Universit\"at Heidelberg, Zentrum f\"ur Astronomie (ZAH), Institut f\"ur Theoretische Astrophysik, Albert-Ueberle-Str. 2, 69120,\\ $^{\,}$ Heidelberg, Germany\\
    $^{2}$Universit\"at Heidelberg, Interdisziplin\"ares Zentrum f\"ur Wissenschaftliches Rechnen, Heidelberg, Germany\\
    $^{3}$Physics and Astronomy Department Galileo Galilei, University of Padova, Vicolo dell'Osservatorio 3, I--35122, Padova, Italy\\
    $^{4}$INFN, Sezione di Padova, Via Marzolo 8, I--35131 Padova, Italy}

   \date{Received XXX / Accepted YYY}

 
  \abstract{
  Binary black holes (BBHs) forming in the accretion disks of active galactic nuclei (AGNs) represent a promising channel for gravitational-wave production. 
  BBHs are often assumed to form at migration traps, i.e. radial locations where the Type I migration of embedded stellar-mass black holes (BHs) transitions from outwards to inwards. In this work, we test this assumption by explicitly simulating the radial migration of BH pairs in AGN disks under different torque prescriptions, including thermal effects and the switch to Type II migration. We map 
  where and when binaries form as a function of supermassive BH (SMBH) mass, disk viscosity, and migrating BH mass. We find that, for SMBH masses below $\MSMBH^\mathrm{thr} \sim 10^8 \Msun$, the majority of pair-up events occur near migration traps ($\gtrsim 80\%$). In contrast, for higher SMBH masses, differential migration dominates and off-trap pair-ups can prevail. Certain disk configurations (e.g., $\alpha=0.01$, $\MSMBH< 10^{6}\Msun$) present a significant overdensity of pair-ups even in the absence of traps due to traffic-jam accumulations where the gamma profile changes slope sharply. 
  We also investigate hierarchical BBH formation, showing that higher-generation pair-ups cluster more tightly around trap or traffic-jam radii. 
  Our results provide realistic prescriptions for BBH pair-up locations and timescales, highlighting the limitations of assuming fixed BBH formation sites.
  }

   \keywords{black hole physics -- stars: black holes -- stars: kinematics and dynamics -- gravitational waves -- galaxies: active}

   \maketitle
%

\section{Introduction}

The direct detection of gravitational waves (GWs) by the LIGO and Virgo collaborations has revolutionized our ability to study compact object populations in the Universe, most notably binary black holes (BBHs, \citealt{GWTC3_first,GWTC3_second}). As the catalog of GW events continues to grow, a pressing challenge is to identify and characterize the astrophysical environments in which such binaries form and evolve.

One particularly promising environment for BBH formation is the accretion disk of an active galactic nucleus (AGN), where a gas-rich and high-density environment 
provides ideal conditions for dynamical interactions among 
black holes (BHs, e.g., \citealt{McKernan_2012, McFacts1_2024, Bartos_2017, Stone_2017, Yang_2019, Secunda_2020, Tagawa_2020_b, Tagawa_2020, Tagawa_2023, Samsing_2022, Vaccaro_2023, McFacts2_2024, McFacts3_2024, Dittmann_2025, Xue_2025}). 
Indeed, stellar-mass BHs can become embedded in the gaseous disk surrounding a supermassive BH (SMBH, \citealt{wang2023, Rowan_2025b, Whitehead_2025}) and thus interact gravitationally with the disk, experiencing migration torques analogous to those governing planetary migration in protoplanetary disks \citep{Paardekooper_migration, McKernan_2012, Bellovary_2016, Masset_2017, Grishin_2024}.

As BHs migrate radially within the disk, the probability of close encounters and subsequent pair-up into binaries is enhanced \citep{DeLaurentis_2023,Li_2023, Rozner_2023, Rowan_2023,Rowan2_2023,whitehead_2023, Whitehead_2024}. Furthermore, the dissipation of orbital energy through gas drag and GW emission can efficiently harden such binaries, leading to mergers within the AGN's lifetime \citep{Peters_1964, Tiede_2020, Ishibashi_2020, Ishibashi_2024, Rozner_2024, Rowan_2025}.

BBH pair-up in AGN disks can occur either within migration traps, special radial locations where the total torque acting on a BH vanishes, halting its migration and promoting accumulation of several BHs, 
or throughout other regions of the disk, collectively referred to as \textit{the bulk} \citep{Wang_2021}. When two BHs migrate at different rates through the AGN disk, a process we refer to as differential migration, their radial paths may intersect. This increases the likelihood of close encounters that, under the influence of gas drag and GW emission, results in BBH formation. 
\citet{McKernan_2020} find that more than 50$\%$ of BBH mergers originate in the bulk, but hierarchical mergers are efficient primarily within migration traps. Furthermore, \citet{Tagawa_2020} demonstrate that BBHs forming in the bulk through gas dynamical capture tend to migrate toward the migration trap as they harden.

Existing semi-analytical approaches focusing on the hierarchical merger scenario, including the \fastcluster{} population synthesis tool developed by the authors \citep{fastcluster2021, fastcluster2022, Vaccaro_2023, Torniamenti_2024}, assign pair-ups to precomputed trap locations. While computationally convenient, this approach limits the physical realism of model outputs, especially for predicting spatial distributions and formation times of merging binaries.

In this work, we relax this simplifying assumption by explicitly integrating BH migration trajectories within AGN disks. We use the \citet{SG} model to describe the disk structure and employ the publicly available \pagn{} module \citep{Gangardt_2024} to compute self-consistent radial profiles of gas surface density, temperature, and pressure. We evaluate migration torques using both a classical Type I prescription \citep{Paardekooper_migration, Bellovary_2016} and an updated formulation including thermal effects \citep{Masset_2017, Grishin_2024}. We account for transitions to Type II migration under gap-opening conditions \citep{Lin1993, Bryden_1999, McKernan_2012_bis, Vaccaro_2023, Gilbaum_2025}.

We simulate millions of independent BH pairs, drawn from a mass distribution generated with population-synthesis calculations \citep{sevn_2023}, and track their one-dimensional migration under gravitational torques. We account for differential migration rates between BHs of varying masses, and identify binary formation 
using a combination of Hill-stability and orbit-crossing criteria. We further incorporate hierarchical evolution by seeding higher-generation ($Ng$) BHs based on merger outcomes of first-generation ($1g$) binaries, including recoil kicks \citep{maggiore_GW}.

Our main objective is to quantify the spatial distribution of BBH pair-ups across a broad parameter space. This includes variations in SMBH mass, disk viscosity, torque prescription, and BH mass. Our results provide realistic prescriptions for pair-up radii and timescales, which can be incorporated into semi-analytic population synthesis codes to improve merger rate predictions and GW data interpretation.


\begin{figure*}
    \centering
    \includegraphics[width=\linewidth]{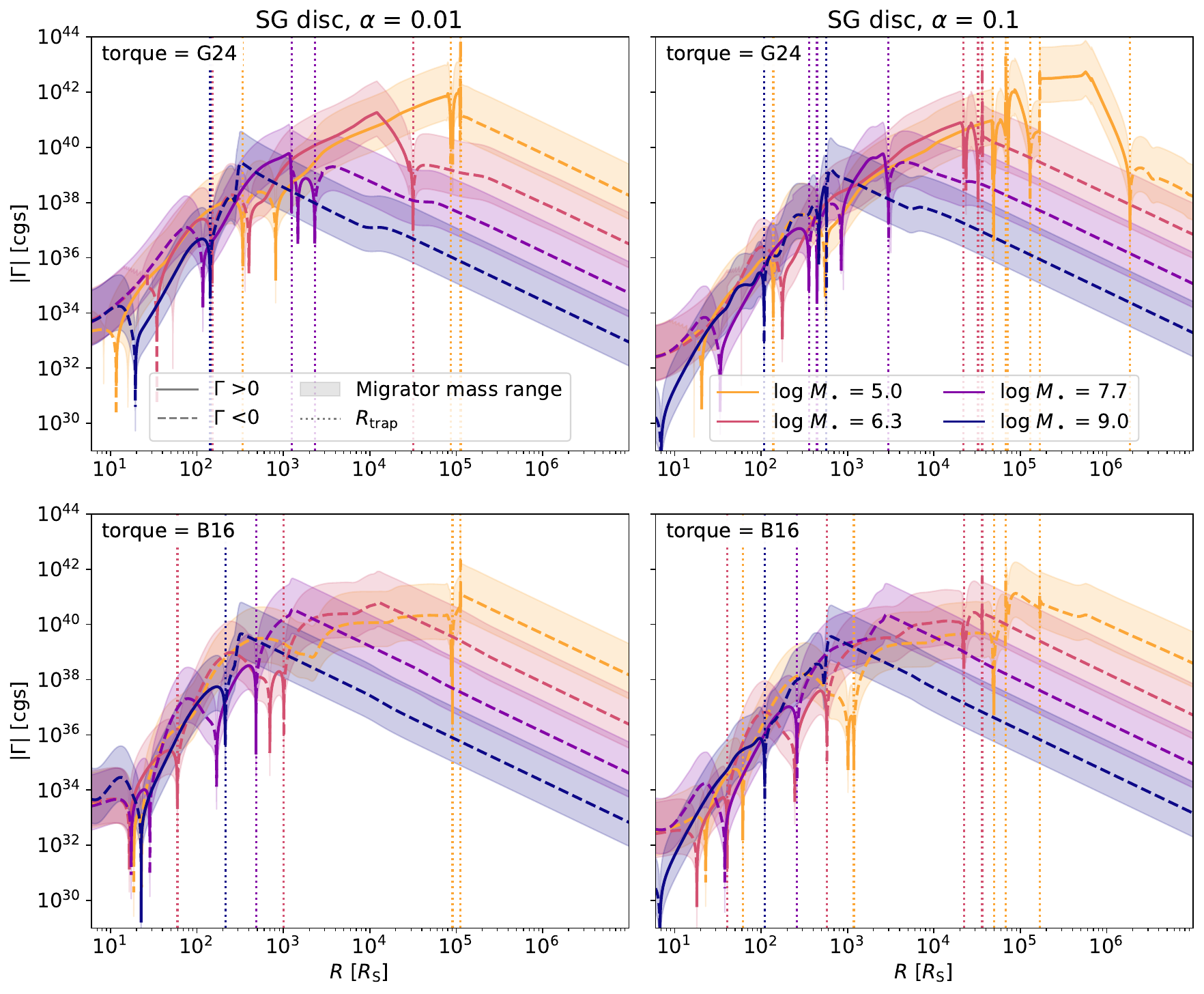}
    \caption{Examples of torque profiles as a function of radial distance (in Schwarzschild radii) for BHs embedded in a \citetalias{SG} disk. Each plot corresponds to a different combination of torque prescription (top: \citetalias{Grishin_2024}; bottom: \citetalias{Bellovary_2016}), 
    disk viscosity parameter $\alpha$ (left: $\alpha = 0.01$; right: $\alpha = 0.1$) and SMBH mass ($\log \MSMBH / \Msun = 5.0, 6.3, 7.7, 9.0$, represented in different colors). Solid and dashed lines indicate positive and negative net torques, respectively, while shaded regions denote uncertainty in the torque determined by computing it over a range of migrator masses between $5 \Msun$ and $50 \Msun$. Vertical dotted lines mark the locations of migration traps.}
    \label{fig:gammas_examples}
\end{figure*}

\begin{figure*}
    \centering
    \includegraphics[width=\linewidth]{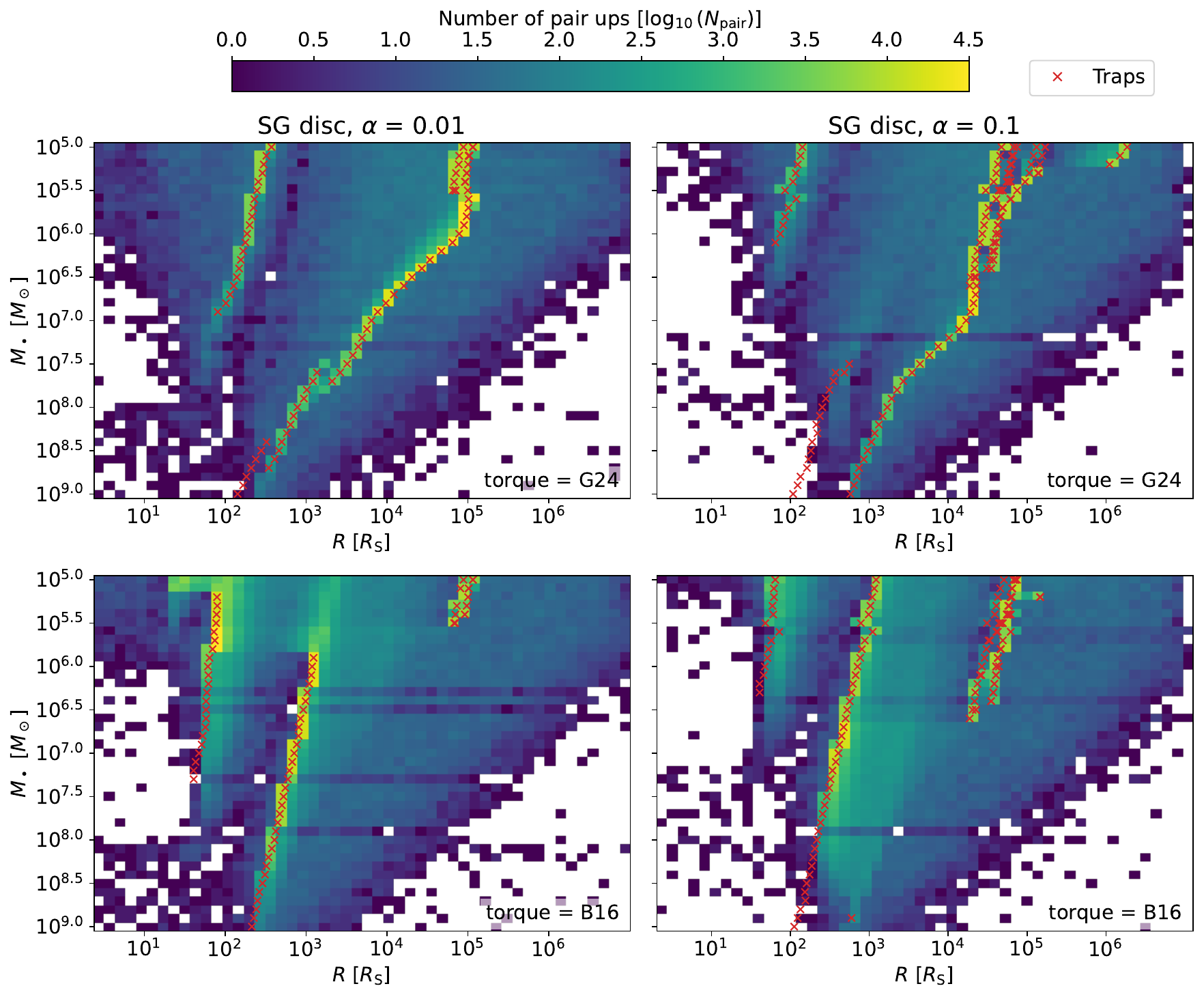}
    \caption{Two-dimensional histograms showing the distribution of BBH pair-up locations in the radius–mass plane ($R$ vs $\MSMBH$), for different combinations of torque prescription (top: \citetalias{Grishin_2024}; bottom: \citetalias{Bellovary_2016}) and disk viscosity $\alpha$ (left: 0.01; right: 0.1) in a \citetalias{SG} disk. The color map indicates the number of pair-ups in each bin $N_\mathrm{pair}$. Red crosses mark the locations of migration traps.}
    \label{fig:catalog_overview}
\end{figure*}

\section{Methods}
\subsection{Disk structure}
We model the physical properties of the accretion disk using the widely accredited steady-state analytic \citealt{SG} disk model (hereafter \citetalias{SG}). 
We solve the corresponding one-dimensional equations self-consistently via the \pagn{} module for the Python programming language \citep{Gangardt_2024} for given input parameters such as the mass of the central SMBH, $\MSMBH$, and the viscosity coefficient, $\alpha$. We fix the mass accretion rate to a tenth of the Eddington rate, $\dot{\MSMBH}=0.1\,\dot{\MSMBH}_\mathrm{,Edd}$. We obtain radial profiles for the gas surface density, $\Sigma_\mathrm{gas}$, the disk aspect ratio, $h=H/R$ (defined as the ratio between the height of the disk, $H\tonde{R}$, and its radius, $R$), the temperature, $T$, and the sound speed, $c_\mathrm{s}$. 

Self-consistent solutions are crucial, as approximate treatments can lead to large deviations in these profiles, which in turn dramatically affect predictions for BBH formation in AGN disks, as discussed in \autoref{sec:appendix_on_vaccaro23_disk}.


\subsection{Migration torques}\label{sec:migration_torques}
We explicitly compute the torques felt by an embedded object of mass $m$ at a radial distance $R$ from a SMBH of mass $\MSMBH$. We include the effects of Type I migration caused by the gravitational perturbation of embedded BHs in the disk \citep{Paardekooper_migration, Lyra_2010, Jimenez_Masset_2017}, accounting also for thermal migration, caused by the thermal response of the disk to the cold, overdense accretion disks surrounding embedded BHs \citep{Masset_2017}. The overall torque is 
\begin{equation}
\label{eq:gamma_tot_typeI}
    \Gamma = \Gamma_{\rm type\, I} + \Gamma_\mathrm{th} 
\end{equation}

We will investigate the difference between a classical Type I migration prescription, inspired by the work by \citealt{Bellovary_2016} (hereafter \citetalias{Bellovary_2016}), and one that also accounts for vertical and thermal effects, similar to \citealt{Grishin_2024} (hereafter \citetalias{Grishin_2024}). 

The classical Type I migration torques are computed according to \citetalias{Bellovary_2016} as
\begin{equation}
    \Gamma_{\rm type\, I} = \frac{\Gamma_\mathrm{ad} \Theta^2 + \Gamma_\mathrm{iso}}{\tonde{\Theta+1}^2} ,
    \label{eq:gamma_type_I}
\end{equation}
where the parameter $\Theta$ is computed according to \citet{Lyra_2010} as\footnote{Here $c_\mathrm{V}$ is the thermodynamic constant with constant volume, $\Omega$ is the Keplerian angular velocity, $\tau_\mathrm{eff}$ is the effective optical depth, and $\sigma$ is the Stefan-Boltzmann constant.}
\begin{equation}
    \Theta=\frac{c_\mathrm{V} \Sigma_\mathrm{g} \Omega \tau_\mathrm{eff}}{12 \pi \sigma T^3} ,
    \label{eq:Theta}
\end{equation}
and the isothermal $\Gamma_\mathrm{iso}$ and adiabatic $\Gamma_\mathrm{ad}$ torques are functions of the adiabatic index $\gamma$ (assumed to be equal to $5/3$ throughout this work) and the gradients of local surface density $\Sigma_\mathrm{g}$, and temperature $T$
\begin{equation}
    a=- \frac{\partial \log \Sigma_\mathrm{g}}{\partial \log R}, \; b=- \frac{\partial \log T}{\partial \log R}, \; c = b - \tonde{\gamma -1}a ;
    \label{eq:disk_gradients}
\end{equation} 
as
\begin{align}
    \label{eq:gamma_type_I_iso}
    \Gamma_\mathrm{iso} / \Gamma_0 =& -0.85 - a - 0.9 b \\
    \label{eq:gamma_type_I_ad}
    \gamma \Gamma_\mathrm{ad} / \Gamma_0 =& -0.85 - a - 1.7 b +7.9 c/\gamma .
\end{align}

\citet{Jimenez_Masset_2017} incorporate three-dimensional effects in their Type I migration torque,
\begin{equation}
    \Gamma_{\rm type\, I} = \Gamma_\mathrm{L} +\left( \frac{0.46 - 0.96 a + 1.8 b}{\gamma} \right) \,\Gamma_0,
    \label{eq:gamma_typeI_JM17}
\end{equation}
and also accounts for the Lindblad torque $\Gamma_\mathrm{L}$ as
\begin{equation}
\Gamma_\mathrm{L}/\Gamma_0 = (-2.34 +0.1 a -1.5 b) \,f(\chi/\chi_c) ,
\label{eq:Gamma_Lindblad}
\end{equation}
where 
\begin{equation}
   f(x) = \frac{(x/2)^{1/2}+1/\gamma}{(x/2)^{1/2}+1},\; \chi = \frac{16 \gamma \tonde{\gamma -1} \sigma_\mathrm{SB} T^4}{3 \kappa \rho_0^2 H^2 \Omega^2},\; \chi_c=H^2 \Omega .
   \label{eq:params_type_I_M17}
\end{equation}
We take the electron-scattering opacity as $\kappa=0.34\, {\rm cm}^2/{\rm g}$ as in \citet{Gangardt_2024}.

The thermal torques are computed as \citep{Masset_2017, Velasco_Romero_2020}
\begin{equation}
    \Gamma_\mathrm{th} = 1.61 \frac{\gamma -1}{\gamma} \frac{x_c}{\lambda} \tonde{\frac{L}{L_c} \,\frac{4m_c}{m+4m_c} -\frac{2m_c}{m+2m_c}} \frac{\Gamma_0}{h} \tonde{1-e^{-\lambda \tau_V/h}} ,
    \label{eq:Gamma_th}
\end{equation}
where $m_c = c_s \chi/G$ is a critical mass, $\tau_V$ is the vertical optical depth, $x_c$ is the corotational radius of the embedded body and $\lambda$ is the typical size of overdense regions surrounding embedded BHs, evaluated as in \citetalias{Grishin_2024}:
\begin{equation}
    x_c = - \frac{H^2}{3 \gamma R} \frac{\partial \log P}{\partial \log R} , \; \lambda = \sqrt{\frac{2 \chi}{3 \gamma \Omega}}.
\end{equation}

We assume that the embedded BH emits at Eddington luminosity, $L=$ $L_\mathrm{Edd}=$ 4 $\pi G m c/ \kappa$, 
and take $L_c=4\pi Gm \rho\chi/\gamma$ as the critical luminosity \citep{Masset_2017}. 


All migration torques are normalized by
\begin{equation}
    \Gamma_0 = \tonde{\frac{m}{ \MSMBH\, h}}^2 \Sigma_\mathrm{g}\, R^4\, \Omega^2 .
    \label{eq:Gamma_0}
\end{equation}

We identify migration traps as locations in the disk where the total Type I migration torque $\Gamma$ crosses zero, changing sign from positive to negative. These correspond to regions where BHs tend to accumulate. In contrast, locations where $\Gamma$ changes sign from negative to positive are referred to as anti-traps, as they tend to be depleted of BHs.

We also account for the possibility of switching to Type II migration. Namely, we enforce a joint condition on the gas viscosity, migrator mass and local disk thinness, \citep{Crida_2006}
\begin{equation}
    K= \frac{m^2}{\alpha h^5 \MSMBH^2} >11.
    \label{eq:open gap}
\end{equation}

If this condition is met, the migrator opens a gap in the disk and can no longer be subject to the migration torque $\Gamma$. Instead, conservatively assuming that no gas can cross the gap, its migration follows the viscous evolution of the disk's gas, moving inwards on a timescale \citep{McKernan_2012}
\begin{equation}
    t_\mathrm{type\, II} = t_\mathrm{visc} = (\alpha \,{}h^2\,{} \Omega)^{-1}.
    \label{eq:t_visc}
\end{equation}
Our prescription for the disk profile does not allow us to self-consistently open a gap (i.e., set $\Sigma_{\rm gas}\tonde{R}\simeq0$ locally), hence in case of Type II migration we artificially construct the corresponding torque as the rate of change in angular momentum, $L=m R^2 \Omega$,
\begin{equation}
\label{eq:gamma_typeII}
    \Gamma_{\rm type\, II} = \left. \frac{dL}{dt} \right|_{\rm type\, II} \simeq \frac{L}{t_\mathrm{type\, II}} = \alpha m H^2 \Omega^2 .
\end{equation}
We discuss the validity and impact of this assumption in \autoref{sec:appendix_typeII}.

\subsection{Integration scheme and pair-up criterion}
We begin each simulation by selecting two stellar-mass BHs that are assumed to already be embedded in the AGN disk. 
For stellar-origin BHs, henceforth referred to as first generation ($1g$) BHs, masses $m$ are sampled independently from a distribution obtained via population-synthesis modeling \textsc{sevn} for solar metallicity \citep{sevn_2023}. The initial radial location $R_\mathrm{init}$, is drawn from a uniform distribution in logarithmic space. Specifically, the logarithm of the initial radius is uniformly sampled between the logarithm of the disk’s inner radius, assumed to be at the innermost stable circular orbit, $R_\mathrm{min}=6\,{}R_\mathrm{g}$ in this context, and an outer radius $R_\mathrm{max}=\SI{0.1}{\parsec}\,{}(\MSMBH/10^6\Msun)^{1/2}$, beyond which the disk's self-gravity becomes important \citep{Goodman_2003, Yang_2019_bis}. This choice reflects the expectation that BHs can form or become embedded at various locations throughout the AGN disk, and provides a neutral baseline that does not bias our results towards any particular radial scale. However, the stellar-origin BHs we expect to find in an AGN disk come from the surrounding NSC, and the probability that they are captured by the  disk depends on the properties of both the AGN disk and the NSC. As a result, embedded BHs are expected to accumulate preferentially in particular radial ranges, enhancing the probability of binary formation at those locations relative to our log-uniform assumption. 

The subsequent orbital evolution is treated in one dimension, considering only the radial location as a function of time. The equation of motion is derived from angular momentum conservation in a Keplerian potential and takes the form:
\begin{equation}
\dot{R} = \frac{2 \, \Gamma(R)}{m} \sqrt{\frac{R}{G \MSMBH}},
\end{equation}
where $\Gamma(R)$ is obtained as in eq. \ref{eq:gamma_tot_typeI} or \ref{eq:gamma_typeII} depending on whether the conditions for Type II migration (eq. \ref{eq:open gap}) are met. The torque profile is precomputed from the AGN disk model and interpolated at each timestep to reflect the actual mass and location of the migrating body.

We solve this system of coupled differential equations using the second-order Runge--Kutta integration method with embedded third-order error estimation and an adaptive timestep.  
The integration spans the full duration of the AGN disk's activity, which we fix at 2\,Myr, and is initialized with a timestep of $10^3$\,yr.
To capture statistical trends in binary formation, this procedure is repeated for a large number of random initial conditions. We simulate $10^6$ independent realizations, each with different initial masses and orbital radii, resulting in a comprehensive catalog of pair-up events across a wide range of physical parameters.

A central goal of the simulation is to determine whether and when a given pair of migrating single BHs becomes a bound binary. To that end, we randomly select two BHs in our catalog and we label them as primary (mass $m_1$, radial location $R_1$) and secondary ($m_2, R_2$), with the convention that $m_1 \geq m_2$. We apply two complementary criteria for binary formation in post-processing, after completing the orbital integration of each single BH.

The first and most stringent condition is based on Hill stability \citep{Hill_1878}. We compute the instantaneous separation between the two BHs, $d = |R_1 - R_2|$, at each integration step and compare it to their Hill radii. A binary is considered to have formed if the separation falls below the larger of the two Hill radii, hence the two BHs are close enough for their mutual gravitational attraction to overcome tidal forces from the central SMBH:
    \begin{equation}
        d \leq \max_{i=1,2} R_{\mathrm{H}}(m_i, R_i), \qquad
        R_\mathrm{H}(m, R) = R \left( \frac{m}{3(\MSMBH + m)} \right)^{1/3}.
    \end{equation}

To prevent the integrator from skipping over close encounters due to large timesteps or rapid migration, we also monitor the sign of the radial separation $E_{\text{cross}} = R_1 - R_2$. A sign change in $E_{\text{cross}}$ indicates that the BHs have crossed each other's orbits and may have come into sufficiently close proximity to allow for capture. If either the explicit  Hill condition or this indirect orbit-crossing condition is met, a BBH is considered to have formed.

In either case, we record the BBH formation time and its pair-up radius, defined as the mass-weighted average position of the two BHs at the moment of binary formation:
\begin{equation}
R_\mathrm{pair} = \frac{m_1 R_1 + m_2 R_2}{m_1 + m_2}.
\end{equation}

This dual-criterion strategy balances accuracy and robustness, ensuring that bound systems are identified even when the BHs migrate rapidly or interact near the resolution limit of the integrator. However, it neglects the exact binary binding mechanism under the influence of gas torques and disk turbulence, investigated in hydrodynamical simulations \citep[e.g.][]{whitehead_2023}. The resulting catalog of binary formation events provides the foundation for our statistical analysis of BBH assembly in AGN disks.

\subsection{Nth-generation BHs}

For Nth-generation ($Ng$) BHs, which are the remnants of earlier mergers in the AGN disk, we estimate their orbital starting conditions using the outputs of $1g$ simulations. 
Hence, in the $Ng$ simulations, the initial position $R_\mathrm{init}$ of the primary BH is no longer drawn randomly across the disk. Instead, it is sampled directly from the set of $R_\mathrm{pair}$ values obtained in the $1g$ simulation, accounting for the natal velocity kick $v_\mathrm{kick}$ that is imparted during the merger event. The recoil velocity $v_\mathrm{kick}$ is drawn from a precomputed \fastcluster{} catalog \citep{Vaccaro_2023} that already incorporates the dependence on progenitor mass ratio and spin configuration as in \citet{maggiore_GW}. The kick direction is randomized assuming isotropic ejection, and contributions from enclosed gas and NSC members are neglected given their small mass relative to the SMBH. If the kick magnitude exceeds the local escape velocity, the remnant is assumed to leave the system. Otherwise, the object is recaptured and is then able to migrate in the disk following the same torque prescriptions as $1g$ BHs. The shifted radius $R_\mathrm{init}^\mathrm{new}$ is computed by means of simple orbital transfer calculations \citep{Hohmann_1925} as
\begin{equation}
    R_\mathrm{init}^\mathrm{new}=\frac{G \MSMBH}{v_\mathrm{tot}^2} \; \quadre{\frac{R_\mathrm{init}\,v_\mathrm{tot}^2 / G \MSMBH}{2 - R_\mathrm{init}\,v_\mathrm{tot}^2 / G \MSMBH } },
    \label{eq:R_after_kick}
\end{equation}
where $v_{\mathrm{tot}}$ is the vector sum of the circular orbital velocity $v_\mathrm{Kepl} = \sqrt{G \MSMBH / R} $ and the random kick $\vec v_\mathrm{kick}$.

In doing so, we implicitly assume that $1g$ binaries will eventually merge at a radial location not significantly different from where they originally paired up.

We only allow $Ng$ BHs to pair up with first-generation companions (forming $Ng\text{--}1g$ BBHs). This choice is motivated by the low probability of drawing a secondary of generation $M$, given the steeply decreasing $p\tonde{M} \propto 2^{-(M-1)}$ \citep{Zevin-Holz}. Accordingly, the secondary BH in each pair has mass and initial radial position sampled in the same way as in the $1g$ simulations.

This methodology ensures that $Ng$ BHs are seeded in dynamically consistent locations, and allows us to model the spatial memory and redistribution of hierarchical mergers across the AGN disk. Other than the initial radius, the rest of the computation proceeds in an identical way to what previously described for $1g$ BHs. 

\subsection{Description of runs}

We performed a suite of Monte Carlo simulations, running $N=10^6$ realizations for each combination of model parameters. We explore how BBH formation in AGN disks depends on the mass of the central SMBH, the disk’s viscosity, and the underlying migration prescription.

We vary the SMBH mass over the range $\log \MSMBH / \Msun = 5.0 \text{--} 9.0$, encompassing typical values expected across cosmic time \citep[e.g.][]{Greene_2007, Trinca_2022}. 

The disk viscosity for a \citetalias{SG} disk, and more generally for so-called $\alpha$-disk models, is parametrized by coefficient $\alpha \in \quadre{0,\,1}$.
We adopt $\alpha = 0.1$ and $\alpha = 0.4$ to bracket the range suggested by observations of fully ionized accretion disks, which typically yield $\alpha \sim 0.1 - 0.4$ \citep{King_2005, Martin_2019}. 
We also run some models with $\alpha = 0.01$ in order to align with conventional assumptions in theoretical models that often presume lower turbulence efficiency (e.g. standard disk models such as \citetalias{SG}) and magneto-hydrodynamical simulations \citep[e.g.][]{Hawley_2011}. We are limited by the assumption of a spatially constant $\alpha$ across the disk radius, consistent with the approximation in the \citetalias{SG} model, despite more recent simulations indicating potential radial variation in viscosity \citep[e.g.][]{Penna_2012}.

To investigate the impact of different assumptions on the migration mechanism, we distinguish between a \citetalias{Bellovary_2016} prescription, where we set $\Gamma_\mathrm{th} =0$ in eq. \ref{eq:gamma_tot_typeI} and compute $\Gamma_\mathrm{type\,I}$ as in eq.~\ref{eq:gamma_type_I} (as in \citealt{Bellovary_2016}), and a \citetalias{Grishin_2024} prescription where we set $\Gamma_\mathrm{th}$ as in eq. \ref{eq:Gamma_th} and $\Gamma_\mathrm{type\,I}$ as in eq.~\ref{eq:gamma_typeI_JM17} (as in \citealt{Grishin_2024}).

\begin{figure*}
    \centering
    \includegraphics[width=0.9\linewidth]{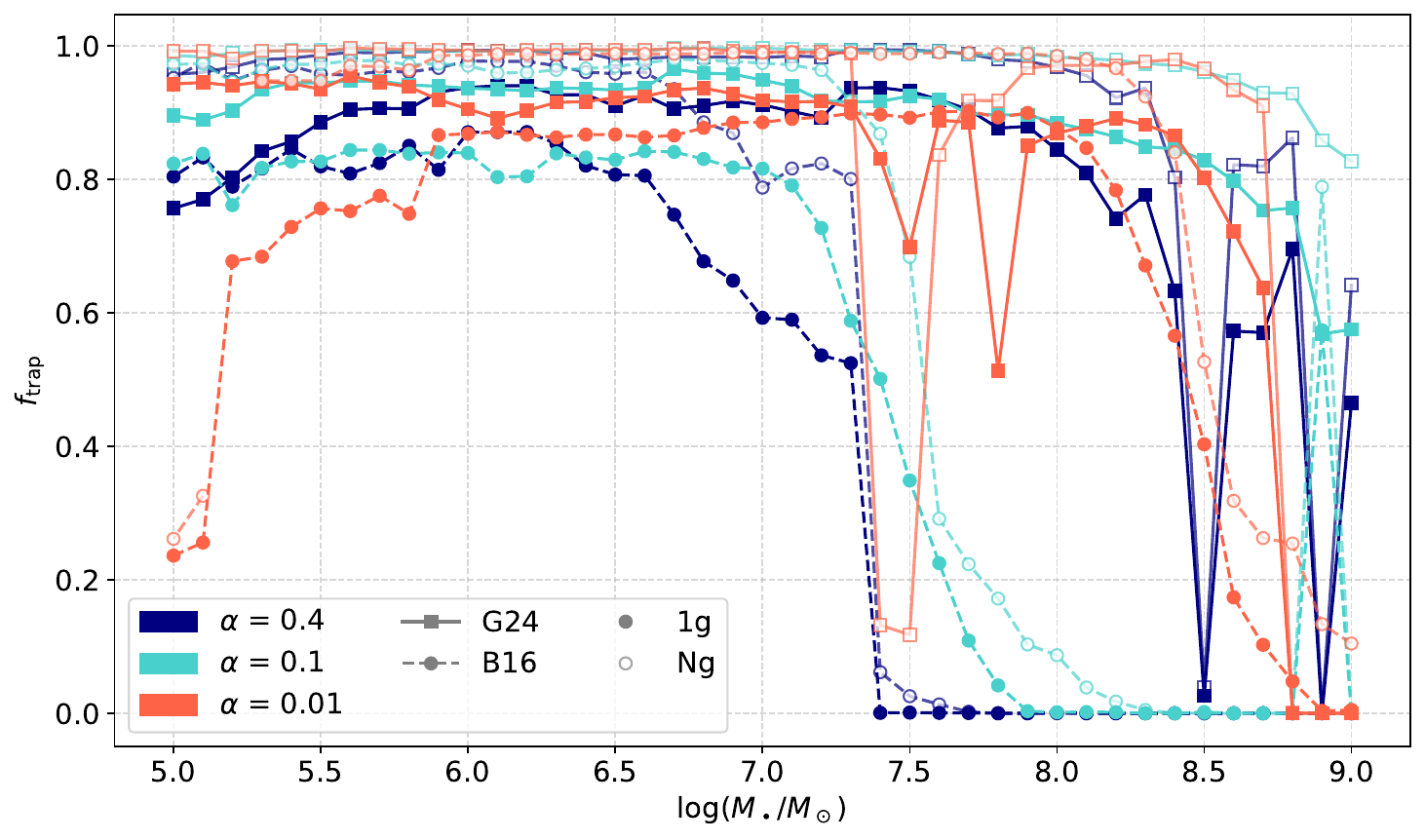}
    \caption{Fraction of binary pair-ups occurring within migration traps as a function of the SMBH mass. The vertical axis shows the occurrence $f_\mathrm{trap}$ of pair-up events that take place in proximity to predefined migration trap locations, while the horizontal axis shows the logarithm of the central SMBH mass. Different marker fillings indicate the BH generation (filled: $1g$; hollow: $Ng$), while color encodes the viscosity parameter $\alpha$ (orange: 0.01; teal: 0.1; navy blue: 0.4). The two torque formalisms are distinguished by marker shape and line style (squares and solid lines: \citetalias{Grishin_2024}; circles and dashed lines: \citetalias{Bellovary_2016}). Trap proximity is defined using a logarithmic tolerance on radial separation (see \autoref{sec:traprelevance} for details).}
    \label{fig:pairup_trap_fraction}
\end{figure*}

\begin{figure*}
    \centering
    \includegraphics[width=\linewidth]{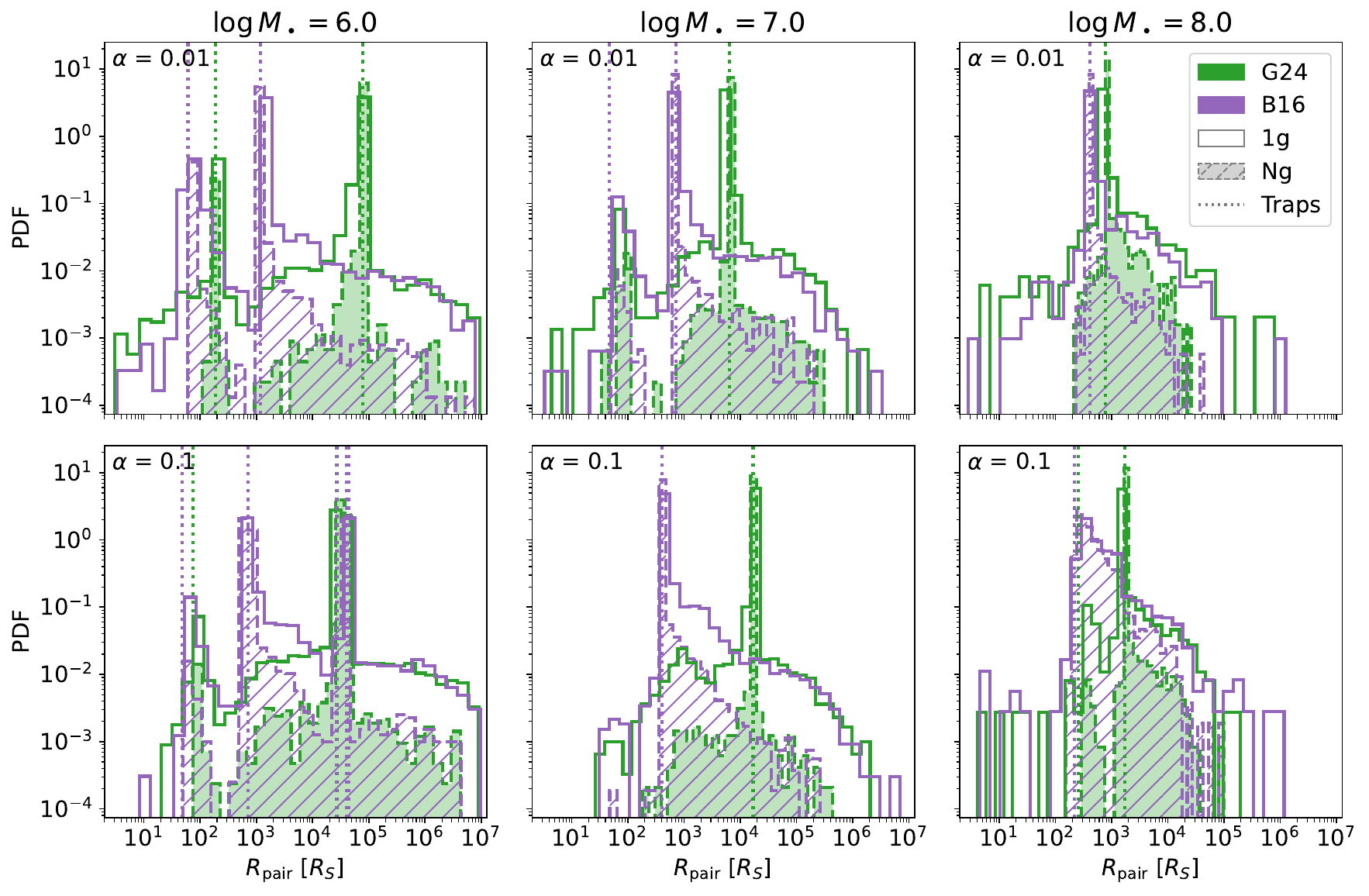}
    \caption{Histograms of binary pair-up radii $R_\mathrm{pair}$ (in units of Schwarzschild radii) for first-generation ($1g$) and $N$th-generation ($Ng$) BHs, across a range of SMBH masses ($\log \MSMBH / \Msun = 6, 7, 8$ respectively in the three columns) and viscosity parameters (upper row: $\alpha=0.01$; lower row: $\alpha=0.1$). Panels show the pair-up radii probability density functions (PDFs), normalized separately in each panel. The $1g$ distributions are shown as step histograms, while $Ng$ distributions are overlaid either as filled or hatched histograms for visual distinction. Vertical dotted lines indicate the positions of migration traps identified from the torque profile. Color represents the migration model used (green: \citetalias{Grishin_2024}; purple: \citetalias{Bellovary_2016}).}
    \label{fig:pairup_1g_vs_Ng}
\end{figure*}

\begin{figure*}
    \centering
    \includegraphics[width=0.9\linewidth]{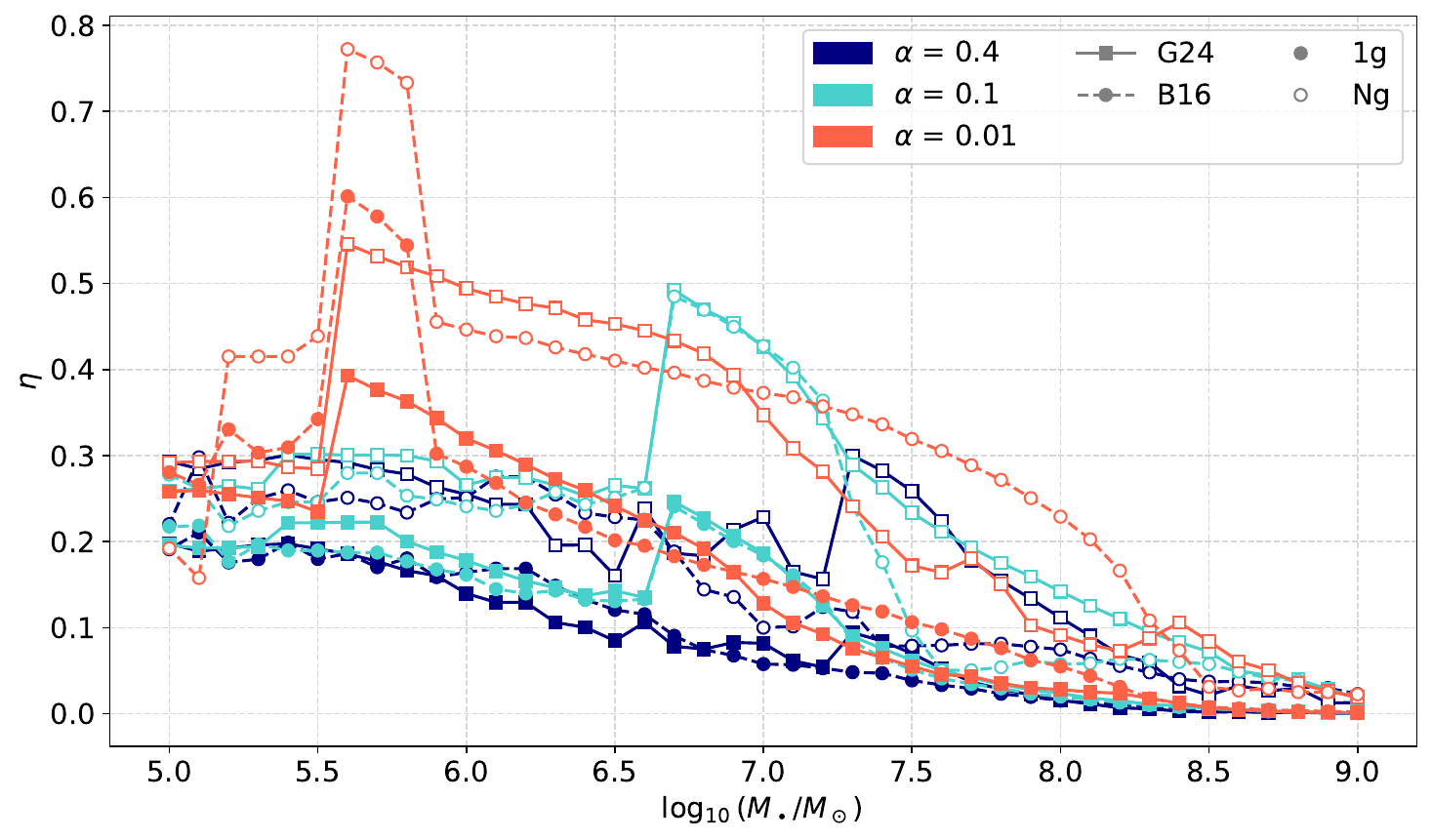}
    \caption{Pair-up efficiency as a function of the central SMBH mass. The vertical axis shows the fraction of BHs that successfully form a binary during the simulation, while the horizontal axis shows the logarithm of the SMBH mass. As in \autoref{fig:pairup_trap_fraction}, different marker fillings indicate the BH generation (filled: $1g$; hollow: $Ng$), color encodes the viscosity parameter $\alpha$ (orange: 0.01; teal: 0.1; navy blue: 0.4), and the two torque prescriptions are distinguished by marker shape and line style (squares and solid lines: \citetalias{Grishin_2024}; circles and dashed lines: \citetalias{Bellovary_2016}).}
    \label{fig:pairup_efficiency}
\end{figure*}

\begin{figure*}
    \centering
    \includegraphics[width=\linewidth]{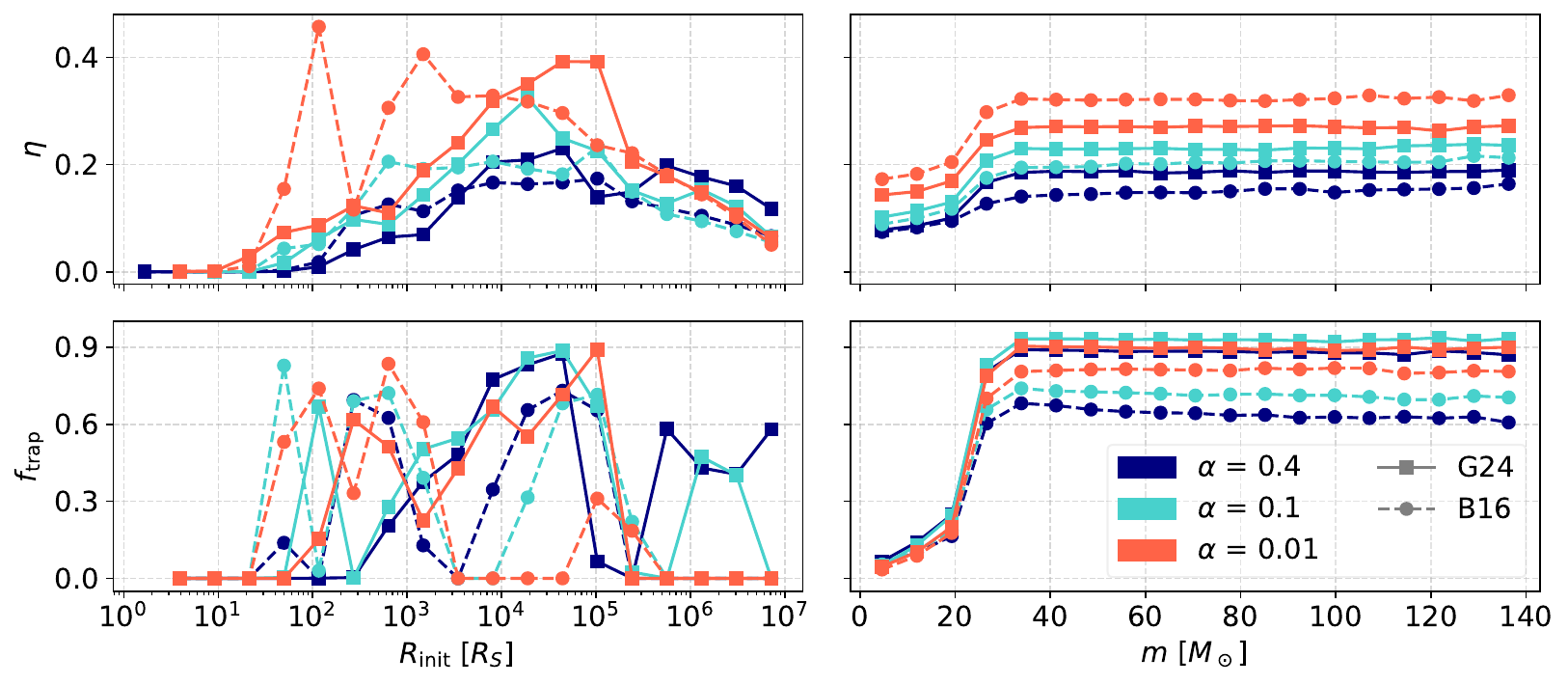}
    \caption{Pair-up efficiency (top panels) and fraction of pair-ups occurring within migration traps (bottom panels) as functions of initial radius $R_\mathrm{init}$ (left) and migrator mass $m$ (right). Color encodes the viscosity parameter $\alpha$ (orange: 0.01; teal: 0.1; navy blue: 0.4), and the two torque prescriptions are distinguished by marker shape and line style (squares and solid lines: \citetalias{Grishin_2024}; circles and dashed lines: \citetalias{Bellovary_2016}).}
    \label{fig:efficiency_trap_fraction_by_r1_m1}
\end{figure*}

\begin{figure*}
    \centering
    \includegraphics[width=\linewidth]{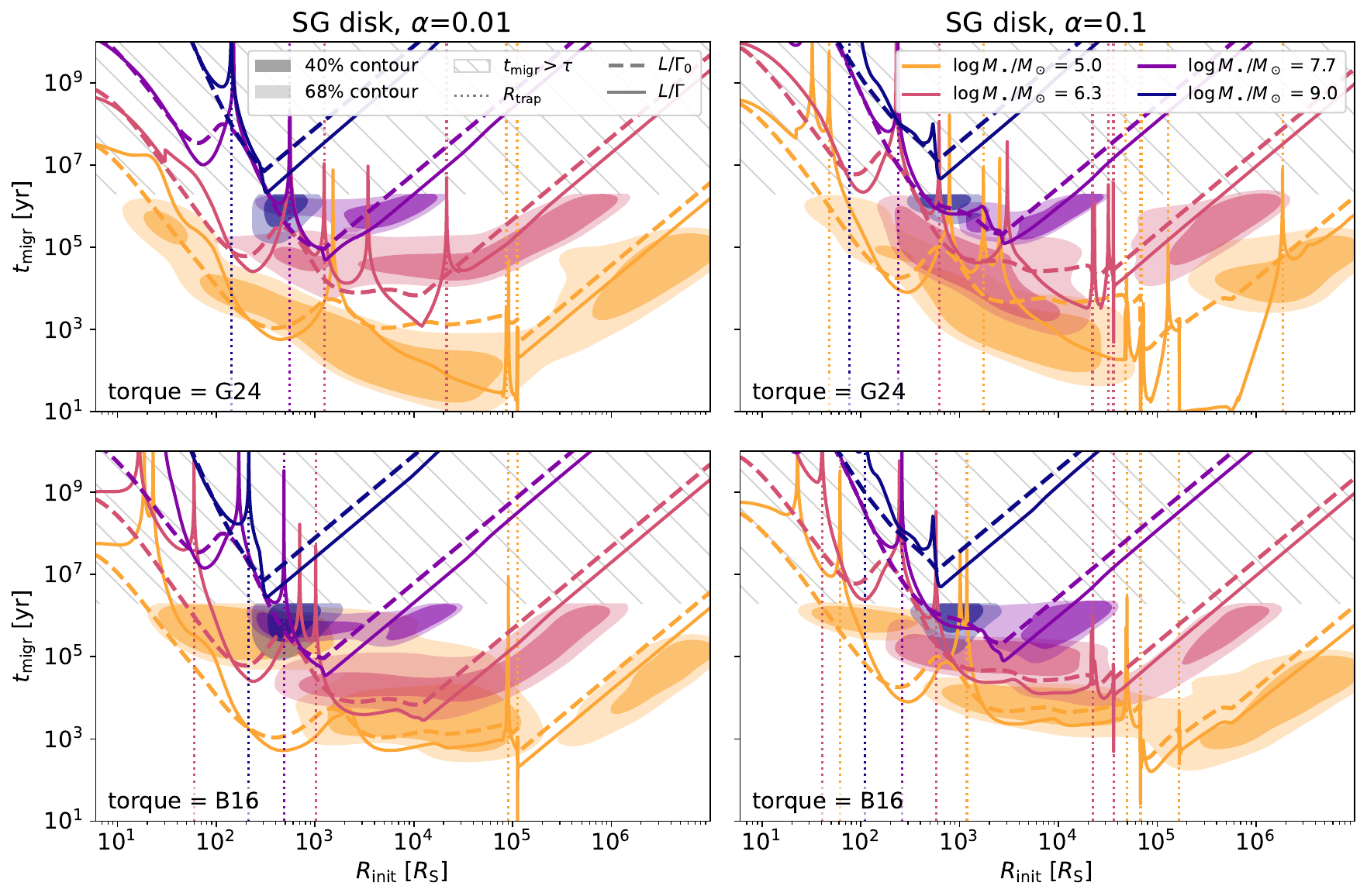}
    \caption{Two-dimensional distributions of the migration time $t_\mathrm{migr}$ as a function of the initial radius $R_\mathrm{init}$ in \citetalias{SG} AGN disks, for different combinations of the viscosity parameter $\alpha$ (left: 0.01; right: 0.1) and torque prescriptions (top: \citetalias{Grishin_2024}; bottom: \citetalias{Bellovary_2016}). Each panel shows contours enclosing $40\%$ and $68\%$ of the migration time distribution, based on our simulations. Results are shown for four representative SMBH masses: $\log(M_{\bullet}/M_{\odot}) = 5.0,\, 6.3,\, 7.7,\, 9.0$, indicated by color. Solid (dashed) lines represent the migration timescale computed using the torque $\Gamma$ in eq. \ref{eq:gamma_tot_typeI} ($\Gamma_0$ in eq. \ref{eq:Gamma_0}). Vertical dotted lines mark the locations of migration traps.}
    \label{fig:timescales_overview}
\end{figure*}

\section{Results}\label{sec:results}
\subsection{Migration torques and pair-up radii}\label{sec:pairup_radii}
We begin by illustrating the torque structure that governs BH migration in \citetalias{SG} disks. \autoref{fig:gammas_examples} shows representative profiles of the total torque acting on a BH as a function of radial distance. 

The torque intensity and sign vary depending on the local properties of the disk such as temperature and density gradients as in eq.s \ref{eq:gamma_type_I} to \ref{eq:params_type_I_M17}, which are themselves functions of $\alpha$, $R$, and $\MSMBH$. The mass of the migrator $m$ only affects the intensity of the torque via the normalization factor in eq. \ref{eq:Gamma_0}. Hence, migration trap locations are disk-specific and independent on the characteristics of the migrating BHs.
Multiple torque sign reversals from $\Gamma >0$ to $\Gamma<0$ are generally present, corresponding to migration traps, whose number and position are of course also sensitive to $\MSMBH$ and $\alpha$. 

To connect this torque structure with the outcome of binary formation, we show in \autoref{fig:catalog_overview} the full distribution of BBH pair-up locations across our catalog of simulations. Each panel displays a two-dimensional histogram in the radius–SMBH mass plane, indicating the logarithmic number of pair-ups in each bin. We also indicate the locations of migration traps, as extracted from the torque profiles. Most BHs that successfully form binaries do so while migrating under Type I migration, with fewer than $8\%$ forming under Type II conditions. 

We find that the radial distribution of pair-ups is broadly consistent with the location of migration traps, supporting the idea that traps are favorable sites for binary formation.
However, the distribution shows a significant spread around traps as well as some clear offsets, with efficient pair-ups also occurring at locations not identified as traps, while some traps contain no BBH pair-ups at all.
These departures can be attributed to differential migration: mismatches in migration speed may cause pair of BHs to pair up
upstream of the trap radius. 

In some cases, such as at $R\simeq10^3 R_\mathrm{S}$ for $\MSMBH<10^6\Msun$ in the \citetalias{Bellovary_2016} prescription, there is a clear overdensity of pair-ups at locations which are not marked as traps, caused by sharp changes in the slope of $\Gamma$. We refer to these locations as traffic jams.

In systems with smaller SMBHs ($\log \MSMBH/\Msun \lesssim 6$), pair-ups are possible at almost all radial locations in the disk, whereas at higher SMBH masses ($\log \MSMBH/\Msun \gtrsim 7$), the distribution of pair-up radii becomes significantly narrower and more concentrated toward the inner disk regions ($10^2 \lesssim R/R_\mathrm{S} \lesssim 10^4$). This narrowing reflects the increased gravitational influence of more massive SMBHs which dominates over local disk perturbations, leading to a more stable torque structure. In particular, the radial torque $\Gamma$ at large radii is less intense and uniformly negative at high $\MSMBH$, reducing differential migration between BHs and suppressing their pairing in the outer disk region.

We observe relevant differences in the trap location based on the torque prescription, as shown in \autoref{fig:catalog_overview}. Using the full \citetalias{Grishin_2024} migration torque gives rise to outer migration traps which are absent in the classical \citetalias{Bellovary_2016} case for SMBH masses $\log \MSMBH/\Msun \in [5.6, \sim8.0]$ ($[6.7, 9.0]$) for $\alpha=0.01$ ($0.1$).

\subsection{The relevance of migration traps}\label{sec:traprelevance}

To quantify the role of migration traps in BBH formation, we track the radial locations at which BHs pair up and compare them to the positions of predefined migration traps in the disk. For each simulation, we identify a set of trap locations $\{R_\mathrm{trap}\}$ based on the local torque profile, as described in \autoref{sec:migration_torques}. A pair-up event is considered to occur ``within a trap'' if the radial distance between the pair-up location $R_\mathrm{pair}$ and any trap satisfies the following logarithmic tolerance condition:
\begin{equation}
    \left| \log R_\mathrm{pair} - \log R_\mathrm{trap} \right| < \delta,
    \label{eq:log-tolerance}
\end{equation}
where $\delta$ is a fixed tolerance. In this work, we adopt $\delta = 0.05$.

Based on this definition, we compute the fraction $f_\mathrm{trap}$ as
\begin{equation}
    f_\mathrm{trap} = \frac{N_\mathrm{in\,trap}}{N_\mathrm{pair}},
\end{equation}
where $N_\mathrm{in\,trap}$ is the number of pair-up events satisfying the condition in eq.~\ref{eq:log-tolerance}, and $N_\mathrm{pair}$ is the total number of pair-up events recorded in the simulation. 

Figure~\ref{fig:pairup_trap_fraction} shows the resulting fraction $f_\mathrm{trap}$, plotted as a function of the central SMBH mass.
Overall, the majority of pair-ups occur within migration traps across most of the parameter space explored. 

For SMBH masses below a certain threshold $M_\bullet^\mathrm{thr}$, $f_\mathrm{trap}$ has values of $0.7$ to $0.9$ for $1g$ BHs and it approaches unity for $Ng$ BHs.
As the SMBH mass increases, $f_\mathrm{trap}$ presents a pronounced transition from values close to unity to nearly zero, indicating that pair-up events are 
no longer 
connected with trap locations at high SMBH masses. This transition typically takes the form of an abrupt drop from $f_\mathrm{trap}\simeq1$ to $f_\mathrm{trap}\simeq0$, with specific value of $M_\bullet^\mathrm{thr}$ depending on the disk and torque model adopted. 

In the high-viscosity \citetalias{Bellovary_2016} models ($\alpha = 0.1, 0.4$), the drop occurs early, at $M_\bullet^\mathrm{thr} \simeq 10^{7.4}\Msun$, whereas in the low-viscosity case ($\alpha = 0.01$), $M_\bullet^\mathrm{thr} \simeq 10^{8.5}\Msun$. For the \citetalias{Grishin_2024} model, the transition to low $f_\mathrm{trap}$ consistently occurs at large SMBH masses; specifically, at $M_\bullet^\mathrm{thr} \simeq 10^{8.8}\Msun, > 10^{9}\Msun, \simeq 10^{8.5}\Msun$ for the cases with $\alpha = 0.01, 0.1, 0.4$.

Furthermore, $f_\mathrm{trap}$ curves present some outliers due to peculiar disk properties for specific combinations of $\MSMBH$ and $\alpha$ (see \autoref{fig:catalog_overview} and additional $|\Gamma|$ plots in the supplementary material). For instance, in disks with $\log \MSMBH/\Msun \in \left[5.0, 6.0\right]$ and $\alpha=0.01$, $f_\mathrm{trap}$ increases gradually from $0.2$ to ${\sim0.8}$ for the \citetalias{Bellovary_2016} torque due to traffic jams: a sharp transition in the slope of $\Gamma$ gives rise to many pair-up events outside of traps due to differential migration. Specifically, as visible in \autoref{fig:gammas_examples}, this happens at $R\lesssim 10^2 \,{}R_\mathrm{S}$ for $\log \MSMBH/\Msun=5.0\text{--}5.1$ and at $R\sim 10^3\,{} R_\mathrm{S}$ for $\log \MSMBH/\Msun<6.0$. 

We show normalized histograms of binary pair-up radii $R_\mathrm{pair}$, split by BH generation ($1g$ vs $Ng$), in \autoref{fig:pairup_1g_vs_Ng} for selected SMBH masses and disk viscosity. Pair-up locations for $Ng$ BHs are considerably more concentrated around peaks than for $1g$ BHs. This enhanced localization arises from the initial conditions: in our model, $Ng$ remnants resume migration at a radius set by Eq.~\ref{eq:R_after_kick}. Because recoil speeds $v_\mathrm{kick}$ are typically small relative to the local Keplerian velocity $v_\mathrm{Kepl}$ \citep{maggiore_GW, Vaccaro_2023}, the kick induces only a minor radial displacement, so $Ng$ remnants remain close to the pair-up radius of their progenitor binary. Since $1g$ binaries preferentially form near peaks in the $R_\mathrm{pair}$ distribution, the ensuing $Ng$ BHs tend to start their migration near those same radii. As a result, $R_\mathrm{pair}$ for $Ng$ objects clusters tightly around peaks, which naturally explains the systematically higher values of $f_\mathrm{trap}$ seen for $Ng$ in \autoref{fig:pairup_trap_fraction} since most peaks are coincident with traps. In contrast, $1g$ BHs are initialized with randomly distributed radial positions within the disk, and their subsequent migration is governed entirely by the local torque profile. While traps still influence the dynamics, the broader range of initial conditions results in a more diffuse distribution of pair-up radii, and therefore a lower $f_\mathrm{trap}$. 

For low viscosity ($\alpha=0.01$) under the \citetalias{Bellovary_2016} prescription, we also find a pronounced pile-up of pair-ups at $R \simeq 10^3 R_\mathrm{S}$ for $\MSMBH \lesssim 10^6 \Msun$ that does not coincide with migration trap locations. This is a direct consequence of differential migration: as shown in \autoref{fig:gammas_examples}, the $|\Gamma(R)|$ profiles exhibit an abrupt change in slope at these radii, producing a traffic-jam accumulation of pair-ups.

These findings emphasize the importance of migration traps as anchoring points for BBH formation in AGN disks. Their effectiveness is sensitive to both disk properties (through the viscosity parameter $\alpha$) and central SMBH mass, and is particularly crucial for $Ng$ pair-ups and their resulting hierarchical mergers.

\subsection{Pair-up efficiency}\label{sec:pairupefficiency}
We define the pair-up efficiency $\eta$ as the fraction of BHs in a given simulation that successfully form a binary by the end of the AGN disk lifetime. That is,
\begin{equation}
    \eta = \frac{N_\mathrm{pair}}{N_\mathrm{tot}},
\end{equation}
where $N_\mathrm{pair}$ is the number of BHs that participated in a pair-up event, and $N_\mathrm{tot}$ is the total number of simulated BHs. \autoref{fig:pairup_efficiency} shows the behavior of $\eta$ as a function of the central SMBH mass, for both $1g$ and $Ng$ BHs. As expected, the efficiency tends to decrease with increasing SMBH mass, due to the suppression of differential migration in the outer regions of the disk discussed in \autoref{sec:pairup_radii}. 

Nonetheless, $Ng$ BHs consistently exhibit higher pair-up efficiency than their $1g$ analogues. This is a consequence of their more favorable initial conditions as previously observed for $f_\mathrm{trap}$: they typically start closer to trap or traffic-jam locations, which increases their likelihood of encountering a suitable companion before the end of the disk lifetime.

We also observe a systematic, although modest, dependence of $\eta$ on the value of $\alpha$ and on the assumed migration model. For low-viscosity disks ($\alpha = 0.01$), pair-up efficiency remains higher across a broader range of SMBH masses, whereas higher values of $\alpha$ ($0.1, 0.4$) slightly reduce the likelihood of BBH formation. Interestingly, these differences are more pronounced for the $Ng$ BHs than for $1g$, where the impact of $\alpha$ is comparatively marginal. 

This may be due to the inherently stochastic nature of pair-up dynamics in the outer disk regions at high viscosity ($\alpha \geq 0.1$), relatively low SMBH masses ($\log \MSMBH / \Msun \lesssim 6.5$), where multiple nearby trap locations (noticeable in \autoref{fig:gammas_examples} and \autoref{fig:catalog_overview}) can lead to  
complex migration patterns. 

We also note marked jumps in $\eta$ which correlate with the temporary disappearance of outer migration traps in the corresponding $\MSMBH$ range, which allows outer BHs to move towards intermediate-radius ($\sim 10^3 R_\mathrm{S}$) locations, characterized by sudden changes in the slope of $\Gamma$ which facilitate BBH captures by differential migration. This happens most notably at $\log \MSMBH / \Msun \sim 5.6\text{–}5.8$ for $\alpha = 0.01$, and at $\log \MSMBH / \Msun \gtrsim 6.7$ for $\alpha = 0.1$.

\subsection{Impact of initial conditions}
\label{sec:parameters_in_results}
\autoref{fig:efficiency_trap_fraction_by_r1_m1} is an analysis of how the pair-up efficiency $\eta$ and the trap fraction $f_\mathrm{trap}$ depend on the initial conditions of the migrating BH, specifically its initial radial position $R_\mathrm{init}$ and mass $m$.

The pair-up efficiency $\eta$ shows a strong radial dependence, peaking at intermediate radii ($R_\mathrm{init} \sim 10^2\text{--}10^5 R_\mathrm{S}$). This trend reflects the underlying structure of the torque profile: migration is inefficient near the inner and outer edges of the disk, while the central regions allow for efficient dynamical pairing. In the innermost disk, pair-ups are suppressed by limited radial volume and weaker torques, whereas in the outer disk, BHs typically migrate too slowly to encounter a companion within the AGN lifetime. A secondary dependence on mass is also evident: heavier BHs ($m \gtrsim 20 \Msun$) tend to achieve higher pair-up efficiencies due to their stronger interaction with the gas and faster migration rates.

The trap fraction $f_\mathrm{trap}$ follows a similar, but more pronounced, trend to the pair-up efficiency $\eta$ in terms of BH mass $m$. In particular, stellar-mass BHs with $m \lesssim 20 \Msun$ are generally unable to reach migration traps, with average $f_\mathrm{trap}$ remaining below $0.3$, whereas BHs with $m \gtrsim 20 \Msun$ consistently exhibit trap fractions exceeding $0.6$ ($0.9$) in the \citetalias{Bellovary_2016} (\citetalias{Grishin_2024}), indicating that higher-mass objects are not only more efficient migrators but also more likely to pair up in the vicinity of traps.

However, the dependence of $f_\mathrm{trap}$ on the initial radius $R_\mathrm{init}$ appears more complex than that of the overall efficiency $\eta$. While the peak values of both quantities occur broadly in the same intermediate radial range ($\sim 10^2\text{--}10^5 R_\mathrm{S}$), the trap fraction displays sharper and more localized peaks. This suggests that proximity to a trap at birth plays a significant role: BHs that happen to become embedded sufficiently close to a stable trap are more likely to remain in its vicinity and undergo in-trap pair-up. Consequently, $f_\mathrm{trap}$ is subject to a degree of stochasticity tied to the initial orbital configuration, which is less evident in the smoother behavior of $\eta$.

\subsection{Migration timescales} 

In semi-analytical models of BH dynamics in AGN disks, the migration timescale is often estimated using the rate of angular momentum change via gaseous torques. A common approximation \citep[e.g.][]{McKernan_2012, Vaccaro_2023} is to compute the migration time as the ratio between the angular momentum $L$ of the BH and the torque normalization $\Gamma_0$:
\begin{equation}
    t_{\mathrm{migr},0} = \frac{L}{\Gamma_0} = \frac{\MSMBH^2 h^2}{m \Sigma_\mathrm{g} R^2 \Omega}
\end{equation}
where $L = m R^2 \Omega$ and $\Gamma_0$ is as in eq. \ref{eq:Gamma_0}. This simplified expression is widely used in the literature as it captures the characteristic timescale over which a BH would drift through the disk under idealized conditions. In \autoref{fig:timescales_overview}, the resulting timescales computed for a $10 \Msun$ BH are shown as dashed lines for different SMBH masses, viscosity parameters, and torque prescriptions.

However, when using the full torque $\Gamma$ (eq. \ref{eq:gamma_tot_typeI}) the corresponding migration timescale,
$t_\mathrm{migr} = L / \Gamma$,
deviates from the $\Gamma_0$-based estimate, especially in regions near torque sign-reversal points, i.e., in the vicinity of migration traps or anti-traps, where $\Gamma=0$. As a result, the solid lines in \autoref{fig:timescales_overview} exhibit sharp features and even divergences, reflecting the breakdown of the timescale approximation in those zones.

To assess how well either of these approximations tracks the actual dynamical evolution of BHs in the disk, we compare them to the pair-up times, $t_\mathrm{pair}$, obtained directly from our simulations. These pair-up times, shown via the $40\%$ and $68\%$ contour levels in \autoref{fig:timescales_overview}, represent the elapsed time between the moment at which the single primary BH becomes embedded in the disk and the subsequent formation of a BBH. Unlike $t_\mathrm{migr}$, they are measured durations rather than derived timescales, and naturally include the effects of stochasticity, mass differences, and local disk structure.
Several important features emerge from this comparison. There is substantial scatter in the pair-up times at fixed radius, due in part to the variation in the mass of the migrating BH. There is a sharp upper limit at $t_\mathrm{migr} = 2$ Myr, corresponding to the assumed AGN disk lifetime adopted in our model. Moreover, some radial regions are systematically disfavored for binary formation, as also noticeable in \autoref{fig:efficiency_trap_fraction_by_r1_m1}, due to either negative torques in the innermost regions of the disk or to reduced differential migration in the outer disk regions for large $\MSMBH$, as discussed in \autoref{sec:pairup_radii}. Finally, pair-up times in our simulations are typically shorter than the estimated migration timescale in regions where the $\Gamma$ profile has a large slope, and longer where the profile is flatter. This is a consequence of differential migration: pair-ups are set by relative drift rather than absolute torque. Where the migration speed profile is steep, nearby objects converge quickly and pair up fast. Vice versa, where the profile is flat, velocity differences are only mass-dependent and consequently pair-up times lengthen.

\section{Caveats} 
Our current approach still presents several limitations.

First, we do not model gas accretion onto individual BHs. Accretion is expected to lead to spin alignment and mass growth, which in turn influences migration rates and pair-up probabilities. 
We also neglect any vertical stratification in the AGN disk and restrict all BHs to move strictly in the midplane, thereby omitting any effect related to vertical motion. In addition, the \citetalias{SG} model is not fully realistic, as it assumes an unspecified heating source and does not account for gas consumption even in gravitationally unstable regions.

Our treatment of Type II migration is deliberately conservative, as gap-opening objects are assumed to effectively prevent any inflow across the gap. More realistically, the transition to Type II is not a hard switch: the gap would not open fully and gas inflow would not be completely suppressed \citep{Malik_2015}. As a result, the Type II regime would more closely resemble a modified Type I migration with reduced local density (as shown in \citealt{Kanagawa_2018, Fung_2014}). 
We explore this alternative model for Type II migration in \autoref{sec:appendix_typeII}.

Moreover, our model does not include resonant interactions, such as mean-motion resonances, which could alter or halt migration \citep{EpsteinMartin_2025, Moncreiff_2025}. We also do not consider dynamical formation via three-body scatterings, which is another viable mechanism of BBH formation in AGN environments \citep{Tagawa_2020}. 
Finally, while our pair-up criterion identifies candidate binaries based on proximity, it does not explicitly model binary binding or survivability under the influence of gas torques and disk turbulence.

We will cure these approximations in a forthcoming study.

\section{Summary}

Semi-analytical models such as \fastcluster{} \citep{fastcluster2021,fastcluster2022,Vaccaro_2023, Torniamenti_2024} and \mcfacts{} \citep{McFacts1_2024,McFacts2_2024,McFacts3_2024} have proven to be useful tools for modeling hierarchical mergers of BHs in AGN disks. Their computational efficiency allows for rapid exploration of the parameter space and facilitates comparisons with observational data. However, these models often rely on simplified assumptions regarding the locations at which BBHs form. A common assumption, made for example by \citealt{Vaccaro_2023}, is that all pair-ups occur at so-called migration traps. While this assumption captures part of the physical picture, it overlooks the diversity of pair-up locations that can arise from realistic differential migration.

In this work, we have achieved a more realistic distribution of pair-up radii via explicit integration of migration torques computed for self-consistent \citetalias{SG} AGN disk models modeled with the software \pagn{} \citep{Gangardt_2024}. We identify migration traps as any location in the disk where the torque changes sign from positive to negative. Our key findings are as follows:
\begin{itemize}
    \item The radial distribution of BBH pair-ups mainly follows the locations of migration traps, but with some spread and offsets due to differential migration. The overall pair-up radii profile is mostly influenced by the mass of the SMBH $\MSMBH$, specifically:
        \textit{(i)} For $\log \MSMBH / \Msun \lesssim 6.5$, BBH pair-ups occur broadly across the entire disk; 
        \textit{(ii)} for $\log \MSMBH / \Msun \gtrsim 6.5$, pair-ups concentrate in the inner disk ($10^2 \lesssim R/R_\mathrm{S} \lesssim 10^4$).
    \item The efficiency of BBH formation $\eta$ is mildly influenced by the disk parameters, decreasing for increasing $\MSMBH$ and $\alpha$, and it it mostly influenced by the initial conditions of the migrating BHs. Indeed, BBHs are formed more efficiently for intermediate initial radii ($R_\mathrm{init}\sim 10^2-10^5 R_\mathrm{S}$) and large migrator mass ($m\gtrsim20 \Msun$).
    \item A large fraction of BBH pair-ups ($f_\mathrm{trap} \gtrsim 80\%$ for first-generation BHs, $f_\mathrm{trap} \sim 100\%$ for $N$-th generation BHs) occur in migration traps for low-to-intermediate SMBH masses 
    $\MSMBH < M_\bullet^\mathrm{thr}$. The threshold value is model dependent: $M_\bullet^\mathrm{thr}\simeq10^{7.4} \Msun$ for the high-viscosity ($\alpha\geq0.1$) in the \citetalias{Bellovary_2016} case, or $M_\bullet^\mathrm{thr}\geq10^{8.5} \Msun$ for all other models. At larger SMBH masses, $f_\mathrm{trap}$ drops to lower values.
    \item For low SMBH masses ($\log \MSMBH/\Msun<6.0$), low viscosity ($\alpha =0.01$) and \citetalias{Bellovary_2016} torque prescriptions, pair-up events pronouncedly pile up at $R\simeq 10^3 R_\mathrm{S}$ as consequence of differential migration: sharp changes in the slope of $|\Gamma(R)|$ create a traffic-jam accumulation even in the absence of a trap.
    \item Estimating migration timescales as the inverse of the rate of angular momentum change ($t_\mathrm{migr}\simeq L/ \dot{L} = L/\Gamma$) tends to overestimate pair-up times in outer disk regions. True pair-up times (from simulations) are shorter due to differential migration and movement into regions of stronger torque. 
\end{itemize}

Our results demonstrate that, although BBH formation predominantly happens in migration traps for a large portion of the explored parameter space, formation in the bulk is non-negligible. 

By integrating the migration trajectories of individual BHs and applying physically motivated pair-up criteria, we recover a distribution of pair-up radii that reflects the full complexity of orbital evolution in AGN disks. This highlights the limitations of using pre-determined locations to represent the entire BBH formation process. 

Incorporating realistic pair-up radius distributions 
into semi-analytic population models will improve the physical accuracy of merger rate predictions. It allows for a more sensible representation of where and when binaries are likely to form, and it captures the contribution of dynamically overtaking encounters which is entirely missed when assuming a static trap location.

\section*{Data availability}
Supplementary material is available via Zenodo at \href{https://zenodo.org/records/16739367}{this link}. 
The data and code used in this study will be made available upon publication.

\begin{acknowledgements}
We thank the anonymous referee for their constructive and insightful comments that improved this manuscript. We thank Kees Dullemond, Dominika Wylezalek, Ralf Klessen, Evgeni Grishin, Nick Stone, Alessandro A. Trani, K.E. Saavik Ford, Barry McKernan, Shawn Ray, Thomas Rometsch, Zoltan Haiman, 
Shmuel Gilbaum, and Lumen Boco for useful discussions. 

MPV and MM acknowledge financial support from the European Research Council for the ERC Consolidator grant DEMOBLACK, under contract no. 770017. They also acknowledge financial support from the German Excellence Strategy via the Heidelberg Cluster of Excellence (EXC 2181 - 390900948) STRUCTURES.  
The authors acknowledge support by the state of Baden-W\"urttemberg through bwHPC and the German Research Foundation (DFG) through grants INST 35/1597-1 FUGG and INST 35/1503-1 FUGG. 

The public version of \textsc{fastcluster} is available via gitlab following \href{https://gitlab.com/micmap/fastcluster_open}{this link}.
We thank Daria Gangardt for developing the \pagn{} software, which is publicly available via github following \href{https://github.com/DariaGangardt/pAGN}{this link}.

This research made use of \textsc{NumPy} \citep{Harris20}, \textsc{SciPy} \citep{SciPy2020}, \textsc{Pandas} \citep{Pandas2024}. Plots were produced using \textsc{Matplotlib} \citep{Hunter2007}.

\end{acknowledgements}

\bibliographystyle{aa} 
\bibliography{bibliography}

\appendix

\section{Improved disk modeling and comparison to previous work} 
\label{sec:appendix_on_vaccaro23_disk}

\begin{figure}
    \centering
    \includegraphics[width=\linewidth]{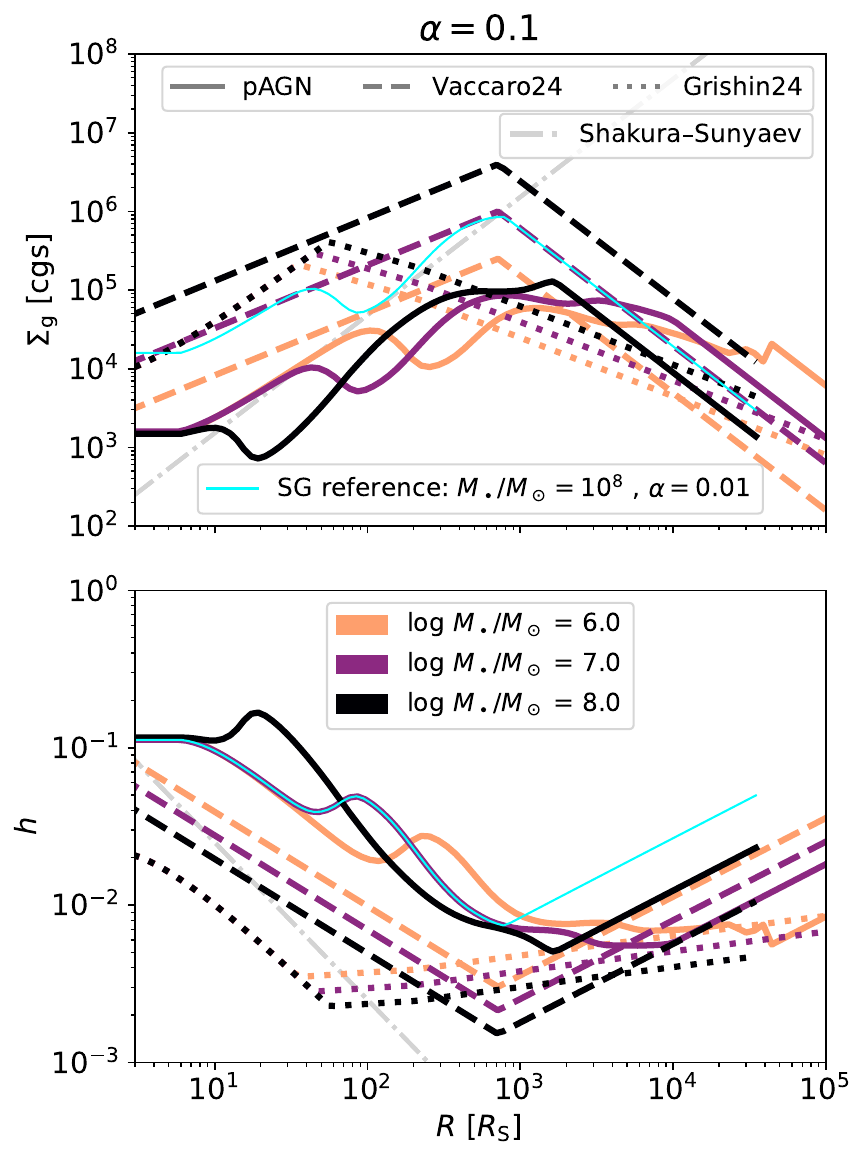}
    \caption{Radial profiles of surface density $\Sigma_\mathrm{g}$ (top panel) and aspect ratio $h=H/R$ (bottom panel) of $\alpha=0.1$ disks, under different numerical treatments. The thin cyan line shows the original \citetalias{SG} profile as published in \citealt{SG}, valid only for $\log \MSMBH / \Msun = 8.0$ and $\alpha = 0.01$. Thicker colorful lines are extensions of the model to different SMBH masses (orange: $\log \MSMBH / \Msun = 6.0$; purple: $7.0$; black: $8.0$). Different line-styles distinguish the numerical method: solid lines are self-consistent solutions of the \citetalias{SG} equations computed with \pagn{}, while dashed and dotted lines represent power-law scaling approximations from \citet{Vaccaro_2023} and \citet{Grishin_2024} respectively. The dot-dashed gray line is a \citet{SSD} disk for comparison. Solid lines should be regarded as the most physically consistent profiles within this framework.}
    \label{fig:SG_disk_models_alpha_0.1}
\end{figure}

\begin{figure}
    \centering
    \includegraphics[width=\linewidth]{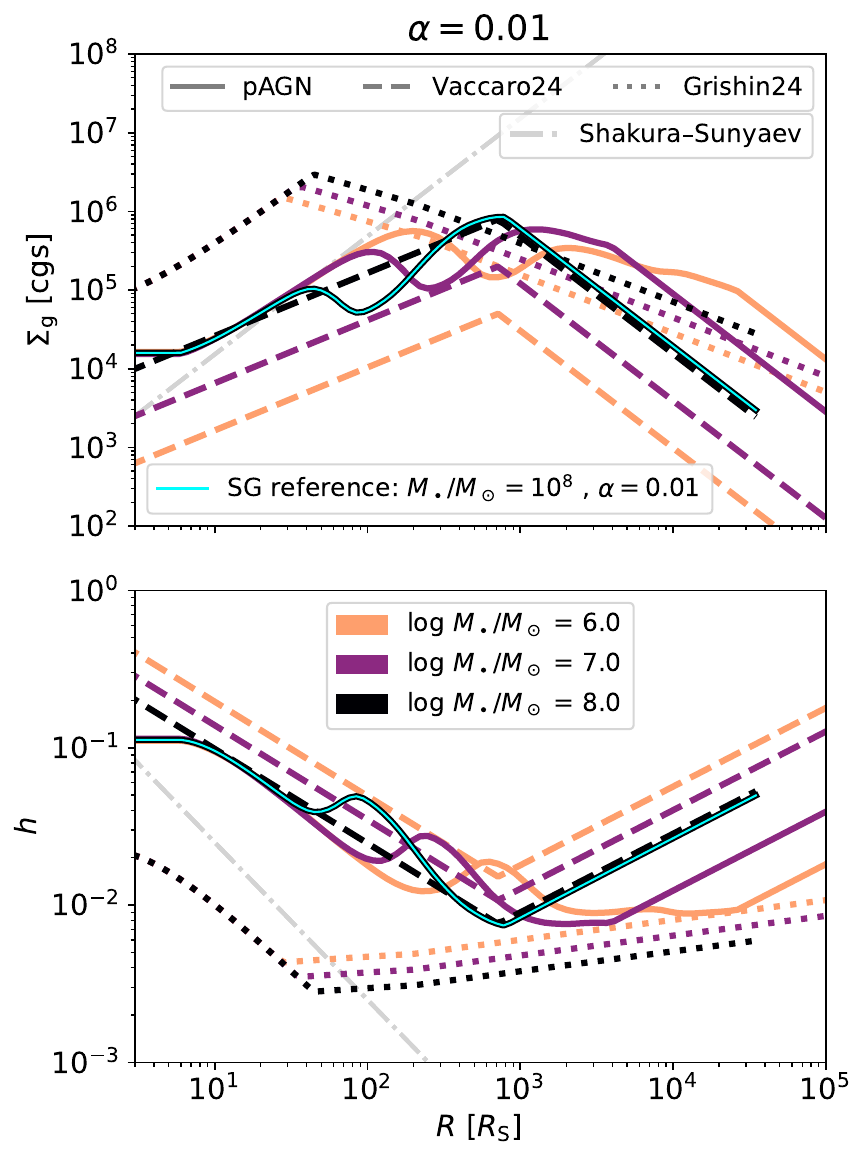}
    \caption{Same as \autoref{fig:SG_disk_models_alpha_0.1}, but for $\alpha = 0.01$.}
    \label{fig:SG_disk_models_alpha_0.01}
\end{figure}

\begin{figure*}
    \centering
    \includegraphics[width=\linewidth]{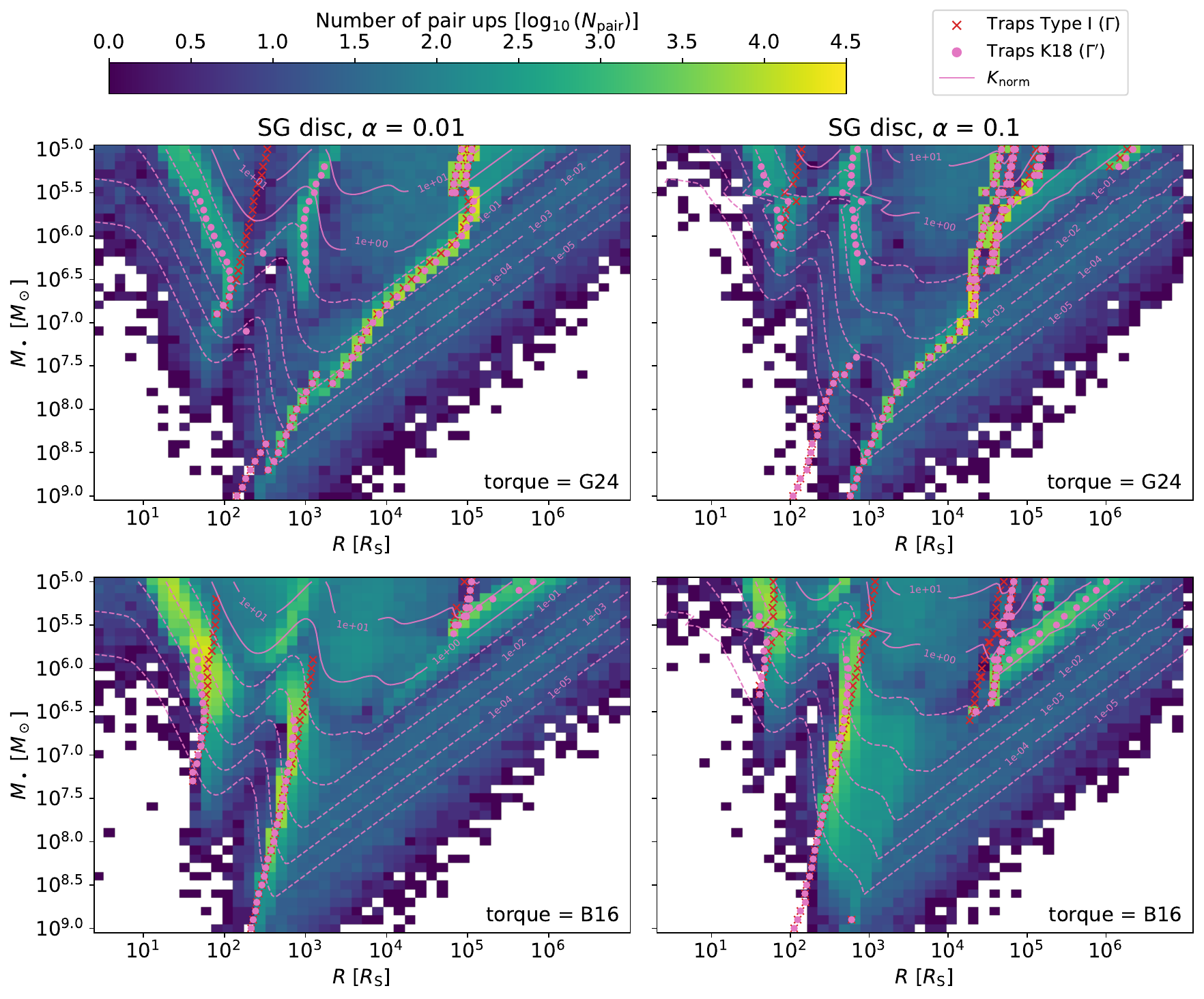}
    \caption{Two-dimensional histograms showing the distribution of BBH pair-up locations in the \citetalias{Kanagawa_2018} model in the radius–mass plane ($R$ vs $\MSMBH$), for different combinations of torque prescription (top: \citetalias{Grishin_2024}; bottom: \citetalias{Bellovary_2016}) and disk viscosity $\alpha$ (left: 0.01; right: 0.1), for a \citetalias{SG} disk model. The color map indicates the number of pair-ups in each bin, $N_\mathrm{pair}$. Pink dots mark the locations of migration traps in the \citetalias{Kanagawa_2018} model, while red crosses mark the Type I trap positions (as in \autoref{fig:catalog_overview}). Pink contours show the values of the normalized contrast parameter $K_\mathrm{norm}$.}
    \label{fig:catalog_overview_K18}
\end{figure*}

A proper and realistic description of the properties of AGN disks is crucial for modeling compact object dynamics in such environments. In this work, we model AGN disks using the analytical tool \pagn{}, developed by \citet{Gangardt_2024}, to numerically solve \citet[][\citetalias{SG}]{SG}  equations of disk structure. A similar approach has recently been adopted in most semi-analytical models for the AGN channel \citep[e.g.][]{McFacts1_2024, McFacts2_2024, McFacts3_2024, Vaccaro_inprep}. This provides smooth profiles for key quantities for any combination of system parameters in a manner fully consistent with the physical description of \citetalias{SG}. The resulting profiles for selected SMBH masses $\MSMBH$ 
are shown in \autoref{fig:SG_disk_models_alpha_0.1} and \autoref{fig:SG_disk_models_alpha_0.01} (hereafter \ref{fig:SG_disk_models_alpha_0.1} and \ref{fig:SG_disk_models_alpha_0.1}) for high ($\alpha=0.1$) and low ( $\alpha=0.01$) viscosity respectively.

In this appendix, we compare our current prescription with  models of $\alpha$-disks previously used in the literature. An $\alpha$-disk is a model in which gas turbulence is 
assumed to be the cause of disk viscosity, characterized by the viscosity coefficient $\alpha \in \quadre{0,\,1}$. 
The prototypical $\alpha$-disk is the one by \citet{SSD}, where crucial quantities such as the gas surface density $\Sigma_\mathrm{g}$ and aspect ratio $h$ 
are modeled as power laws. We show the resulting profile in \ref{fig:SG_disk_models_alpha_0.1} and \ref{fig:SG_disk_models_alpha_0.1}, using the \citet[][eqs.~32--33]{Kocsis_2011} treatment and assuming $\dot{\MSMBH}/\dot{\MSMBH}_\mathrm{,Edd}=0.1$. The obvious problem with this disk model is that becomes unphysical in the outer regions, and it is hence only valid for $R\lesssim 10^3 R_\mathrm{S}$. Moreover, such disks are viscously, thermally, and convectively unstable to perturbations, meaning that they can easily fragment into smaller sub-clouds \citep{Kocsis_2011}.

\citet{SG} extend the \citet{SSD} model by assuming that star formation in the outskirts of the disk acts as stabilizing mechanism against fragmentation. Moreover, the \citetalias{SG} model is also broadly consistent with observed quasar spectral energy distributions, as it more accurately reproduces the flat UV–optical slopes and IR excesses characteristic of luminous quasars. he full model consists of 16 coupled equations describing the radial structure of the disk, which were originally solved self-consistently for a fiducial set of parameters: $\log \MSMBH/\Msun = 8.0$, $\alpha=0.01$, and $\dot{\MSMBH}/\dot{\MSMBH}_\mathrm{Edd}=0.1$. Before the publication of the \pagn{} tool, various approximations were required to generalize this solution to other physical regimes. In the following, we illustrate two examples.

In \citet[eqs.~$1-2$]{Vaccaro_2023}, we fit the original profile with broken power laws, and combine it with a first-order rescaling in $\MSMBH$ and $\alpha$. As shown in \ref{fig:SG_disk_models_alpha_0.1} and \ref{fig:SG_disk_models_alpha_0.1}, this provides quite a crude modeling of the relevant quantities, as it neglects the complex interdependency between parameters in the full system of equations. The approximation breaks down most noticeably at high $\alpha$ and high $\MSMBH$, where it severely overestimates the surface density $\Sigma_\mathrm{g}$ in the inner regions of the disk.

Another approach is the one by \citet[][eqs.~$3-11$]{Grishin_2024} who re-engineer an analytical model which is similar in spirit to \citetalias{SG}, but neglecting some of the physics and using simplified analytical expressions. The resulting profiles, also shown in \ref{fig:SG_disk_models_alpha_0.1} and \ref{fig:SG_disk_models_alpha_0.1} for $\dot{\MSMBH}/\dot{\MSMBH}_\mathrm{Edd}=0.1$, are qualitatively similar to the \citetalias{SG} ones computed with \pagn{} (meaning that they feature an inner region where the density increases and the thickness decreases for growing radii, and an outer region where the opposite happens), but the transition radii are systematically one order of magnitude smaller than in the more accurate \citetalias{SG} paradigm. Moreover, they also 
overestimate the density and underestimate the aspect ratio in the inner regions ($R \lesssim 10^3 R_\mathrm{S}$).

This highlights the importance of solving the full set of equations with \pagn{}, as the dependence of the disk structure on $\alpha$ and $\MSMBH$ is highly nonlinear and cannot be captured by simple rescaling or analytical fits, when constructing models of BH dynamics in AGN disks. 

\section{Type II migration}
\label{sec:appendix_typeII}

\begin{figure*}
    \centering
    \includegraphics[width=\linewidth]{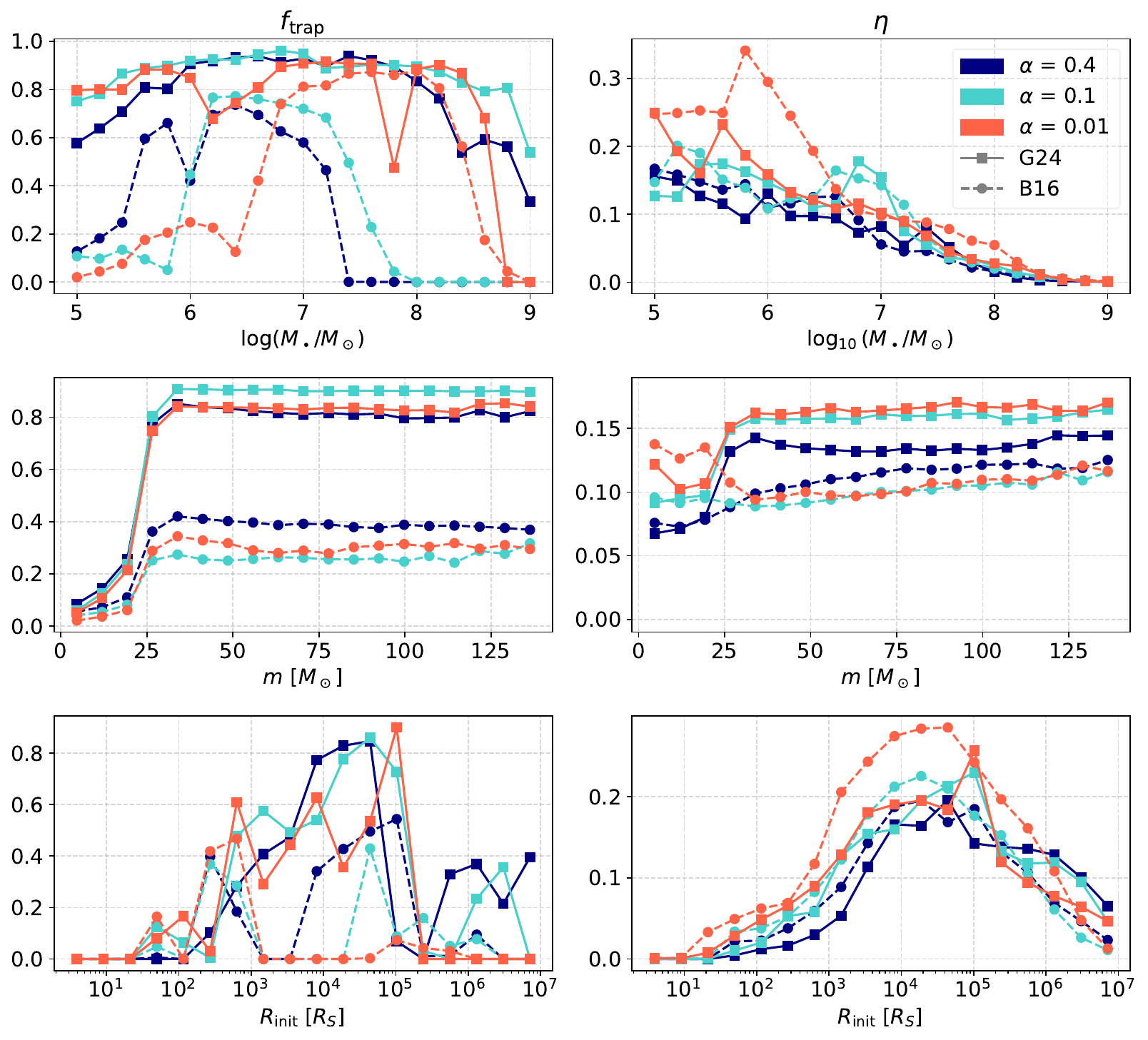}
    \caption{Fraction of pair-ups occurring within migration traps (left panels) and pair-up efficiency (right panels) as functions of SMBH mass $\MSMBH$ (top row), migrator mass $m$ (middle row), and initial radius $R_\mathrm{init}$ (bottom row) in the \citetalias{Kanagawa_2018} prescription. Color encodes the viscosity parameter $\alpha$ (orange: 0.01; teal: 0.1; navy blue: 0.4), and the two torque prescriptions are distinguished by marker shape and line style (squares and solid lines: \citetalias{Grishin_2024}; circles and dashed lines: \citetalias{Bellovary_2016}).}
    \label{fig:efficiency_trap_fraction_K18}
\end{figure*}

\begin{figure*}
    \centering
    \includegraphics[width=\linewidth]{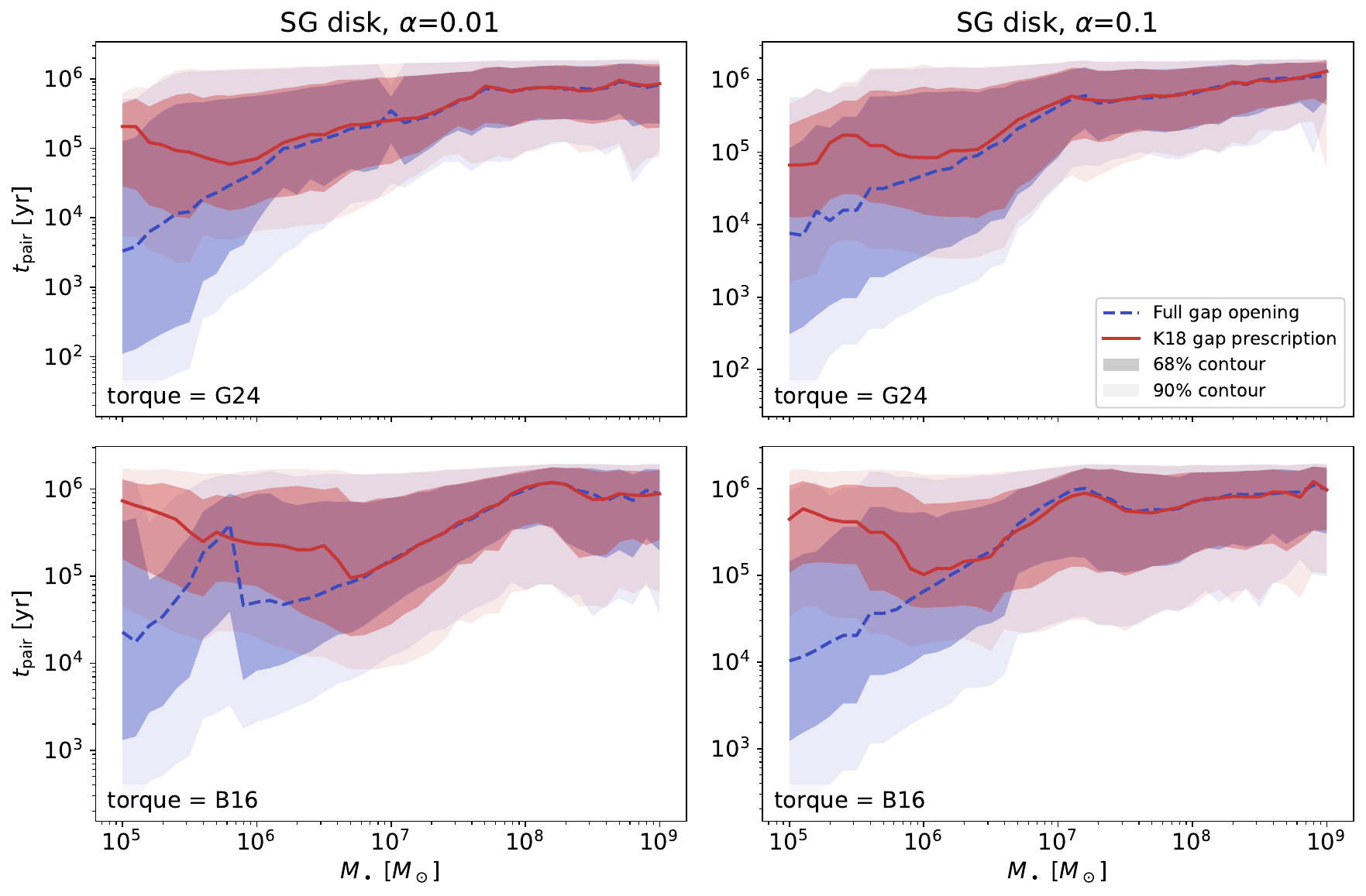}
    \caption{Pair-up times $t_\mathrm{pair}$ as a function of SMBH mass in \citetalias{SG} AGN disks, for different combinations of the viscosity parameter $\alpha$ (left: 0.01; right: 0.1) and torque prescriptions (top: \citetalias{Grishin_2024}; bottom: \citetalias{Bellovary_2016}). Each panel shows a median line and contours enclosing $68\%$ and $90\%$ of the pair-up time distribution, based on our simulations. Results are shown for two different Type II migration prescriptions, indicated by color and line style. In blue, dashed lines: full gap opening, as in the main text (eq.~\ref{eq:gamma_typeII}). In red, continuous lines: partial gap opening, as in this appendix \citepalias{Kanagawa_2018}.}
    \label{fig:timescales_overview_K18}
\end{figure*}

The classical picture of Type II migration assumes that, once a gap is opened, the embedded object becomes locked to the viscous inflow of the disk with migration timescale $t_\mathrm{type\, II} = t_\mathrm{visc}$, as assumed in \autoref{sec:migration_torques}.  
This disk-dominated view is conservative: it assumes the gap to be completely cleared and gas inflow across the gap to be suppressed. Recent hydrodynamical simulations have shown that this picture is oversimplified: in reality, gap opening is often partial and gas can leak through the gap edges, leading to migration rates that are faster than the purely viscous timescale \citep[e.g.][]{Malik_2015}.

\citet{Kanagawa_2018} (hereafter \citetalias{Kanagawa_2018}) proposed a modified prescription that interpolates between Type I and Type II regimes, so that migration is well described as a modified Type I regime with reduced local density
\begin{equation}
    \Sigma_\mathrm{g}^\prime \tonde{R} = \frac{\Sigma_\mathrm{g}\tonde{R}}{1+0.04\,K}, 
\end{equation} 
where the contrast parameter $K$ is defined as in eq.~\ref{eq:open gap}.
For low-mass migrators ($K \lesssim 1$, $\Sigma_\mathrm{g}^\prime \gtrsim 0.96\, \Sigma_\mathrm{g}$), the gap is shallow and standard Type I migration (eq.~\ref{eq:gamma_tot_typeI}) remains valid.
As the migrator mass increases, the contrast parameter $K$ grows quadratically. The full transition to a Type II regime happens for $K \gtrsim 20$ \citepalias{Kanagawa_2018}, as the disk surface density at the orbit is significantly reduced to $\Sigma_\mathrm{g}^\prime \lesssim 0.55\, \Sigma_\mathrm{g}$. 

We calculate corotational gas torques in eq.s~\ref{eq:Theta}, \ref{eq:disk_gradients}, \ref{eq:gamma_type_I_iso}, \ref{eq:gamma_type_I_ad}, \ref{eq:gamma_typeI_JM17}, and \ref{eq:Gamma_0} using the reduced local density $\Sigma_\mathrm{g}^\prime$. In this regime, both the normalization of the torque $\Gamma_0$ (eq.~\ref{eq:Gamma_0}), and the shape of the torque profiles (via the parameter $a$ in eq.~\ref{eq:disk_gradients}) are significantly modified. We leave the Lindblad torque ($\Gamma_\mathrm{L}$, eq.~\ref{eq:Gamma_Lindblad}) unperturbed.

The effect of gap opening of Lindblad torques is quite complex, as it depends on the exact profile of the gap \citep[e.g.][]{Kley_Nelson_2012, Petrovich_2012}. We performed tests in which we rescaled them with the density at the bottom of the gap, $\Sigma^\prime_g$, as we did with corotational torques, with minimal changes to our results. A more comprehensive simulation would include modeling the gap's density profile and computing Lindblad torques self-consistently. However, in the high-viscosity regime in which we are operating, we do not expect any non-linear behavior that would strongly affect our results \citep{Duffell_2015}.

An important implication is that migration traps are no longer determined solely by the background disk structure, as the torques $\Gamma^\prime=\Gamma\tonde{\Sigma_\mathrm{g}^\prime}$ are now $K$-dependent and, consequently, $m$-dependent. This mass dependence adds an additional layer of complexity to the dynamics of BHs in AGN disks. In the following, the trap locations under the \citetalias{Kanagawa_2018} prescription are defined as the radii where the total migration torque $\Gamma^\prime$ acting on a migrator of average mass ($m\simeq13.5 \Msun$) changes sign from positive to negative. 

While the gap contrast naturally depends on the properties of the migrator,a disk-specific contrast parameter $K_\mathrm{norm}$ can be defined by normalizing over the migrator mass:
\begin{equation}
    K_\mathrm{norm}=\frac{K}{m^2}= \alpha^{-1} h^{-5} \left(\frac{\MSMBH}{\Msun}\right)^{-2} 
\end{equation}

\bigskip
We re-compute the binary formation catalog using the \citetalias{Kanagawa_2018} prescription for gap opening. The resulting pair-up distributions are shown in \autoref{fig:catalog_overview_K18}.
Compared to the results presented in \autoref{sec:results}, the locations of migration traps differ substantially, as the modified torques tend to shift the zero-torque radii away from regions of high contrast $K_\mathrm{norm}$.
In the \citetalias{Grishin_2024} scenario, the peaks in the pair-up radius distributions are generally located in proximity to the migration traps, with a few exceptions. In contrast, the \citetalias{Bellovary_2016} case presents a sharp traffic-jam accumulation of pair-up events in valleys of low contrast (low $K_\mathrm{norm}$), even in the absence of a trap.

\autoref{fig:efficiency_trap_fraction_K18} shows the resulting pair-up efficiency, $\eta$, and trap fraction, $f_\mathrm{trap}$, obtained with the \citetalias{Kanagawa_2018} model.

The pair-up efficiency, $\eta$, is generally lower than in the fiducial runs, peaking at roughly $0.3$ for $\MSMBH\lesssim10^6\Msun$ in the low-viscosity \citetalias{Bellovary_2016} case, while retaining similar trends in SMBH mass $\MSMBH$ and in initial radius $R_\mathrm{init}$ to those observed in \autoref{sec:results}. On the other hand, the dependency on migrator mass $m$ is flatter than in the fiducial case, implying that partial gap opening \citepalias{Kanagawa_2018} moderately suppresses the pair-up efficiency of larger mass migrators (with $m\gtrsim20\Msun$).

The fraction of pair-ups occurring within migration traps, $f_\mathrm{trap}$, retains a similar behavior than in \autoref{fig:pairup_trap_fraction} for the \citetalias{Grishin_2024} case, remaining large ($>0.6$) for most of the parameter space, and decreasing for SMBH masses above $10^{8.4} \Msun$. Contrarily, in the \citetalias{Bellovary_2016} case traffic-jam pair-ups are more relevant, especially at low viscosity for low SMBH mass. Therefore, $f_\mathrm{trap}<0.4$ with the exception of a certain $\alpha$-dependent window: $\log \MSMBH/\Msun \in \left[6.6,8.5 \right]$ for $\alpha=0.01$, $\log \MSMBH/\Msun \in \left[6.0, 7.5 \right]$ for $\alpha=0.1$, or $\log \MSMBH/\Msun \in \left[5.5, 7.2 \right]$ for $\alpha=0.4$. It retains a similar trend in both migrator mass $m$ and initial radius $R_\mathrm{init}$ than discussed in \autoref{sec:parameters_in_results}.

We compute pair-up times, $t_\mathrm{pair}$, and evaluate how the \citetalias{Kanagawa_2018} framework affects them. These pair-up times, shown via the $68\%$ and $90\%$ contour levels  in \autoref{fig:timescales_overview_K18}, represent the elapsed time between the moment at which the single primary BH becomes embedded in the disk and the subsequent formation of a BBH. At low SMBH mass, $\MSMBH\lesssim3\times10^6 \Msun$, the normalized contrast parameter can reach larger values, $K_\mathrm{norm}\gtrsim1$, thus significantly reducing the local gas density $\Sigma^\prime$ and slowing down the pair-up process, therefore increasing $t_\mathrm{pair}$ by almost two orders of magnitude for the lowest SMBH mass explored, $\MSMBH=10^5 \Msun$. 
In contrast, for $\MSMBH\gtrsim3\times10^6 \Msun$, pair-up times are unaffected by the adopted Type II prescription.

\bigskip
Overall, the adoption of the \citetalias{Kanagawa_2018} prescription has different consequences depending on the torque model adopted. 

In the \citetalias{Grishin_2024} case, the position of traps is shifted towards regions of low contrast, but both the pair-up efficiency and the relevance of migration traps are mostly unaffected. 

In the \citetalias{Bellovary_2016} case, the formation of migration traps in disks with low SMBH mass ($\log\MSMBH/\Msun\lesssim6$) is fully suppressed, and pair-up events happen in traffic jams at locations of low contrast $K_\mathrm{norm}$. Consequently, the self-consistent treatment of gap opening lowers the pair-up efficiency of high-mass migrators and weakens the correlation between migration traps and binary formation sites, especially in the \citetalias{Bellovary_2016} case.

\end{document}